\documentclass{emulateapj}
\usepackage{lscape}

\newcommand{\lapprox }{{\lower0.8ex\hbox{$\buildrel <\over\sim$}}}
\newcommand{\gapprox }{{\lower0.8ex\hbox{$\buildrel >\over\sim$}}}
\newcommand\iontoo[2]{#1$\;${\scshape{#2}}}

\slugcomment{DRAFT \today}
\shorttitle{PTF10nvg}
\shortauthors{}

\begin{document}

\title{PTF10nvg: An Outbursting Class I Protostar in the Pelican/North American Nebula}

\author{Kevin~R.~Covey\altaffilmark{1,2,3}, Lynne~A.~Hillenbrand\altaffilmark{4}, 
Adam A. Miller\altaffilmark{5}, Dovi Poznanski\altaffilmark{5,7,10},
S. Bradley Cenko\altaffilmark{5},
Jeffrey M. Silverman\altaffilmark{5}
Joshua S. Bloom\altaffilmark{5},
Mansi M. Kasliwal\altaffilmark{4},
William Fischer\altaffilmark{6},
John Rayner\altaffilmark{8,15},
Luisa M. Rebull\altaffilmark{11},
Nathaniel R. Butler\altaffilmark{5,10},
Alexei V. Filippenko\altaffilmark{5},
Nicholas M. Law\altaffilmark{12},
Eran O. Ofek\altaffilmark{10,4},
Marcel Ag\"ueros\altaffilmark{9},
Richard G. Dekany\altaffilmark{13}, 
Gustavo Rahmer\altaffilmark{13}, 
David Hale\altaffilmark{13}, 
Roger Smith\altaffilmark{13}, 
Robert M. Quimby\altaffilmark{4}, 
Peter Nugent\altaffilmark{7}, 
Janet Jacobsen\altaffilmark{7}, 
Jeff Zolkower\altaffilmark{13}, 
Viswa Velur\altaffilmark{13}, 
Richard Walters\altaffilmark{13}, 
John Henning\altaffilmark{13}, 
Khanh Bui\altaffilmark{13}, 
Dan McKenna\altaffilmark{13}, 
Shrinivas R. Kulkarni\altaffilmark{4}, 
Christopher Klein\altaffilmark{5}}

\altaffiltext{1}{Cornell University, Department of Astronomy, 226 Space Sciences Building, Ithaca, NY 14853, USA.}
\altaffiltext{2}{Hubble Fellow.}
\altaffiltext{3}{Visiting Researcher, Department of Astronomy, Boston University, 725 Commonwealth Ave, Boston, MA 02215, USA.}
\altaffiltext{4}{Astrophysics Department, California Institute of Technology, Pasadena, CA 91125, USA.}
\altaffiltext{5}{Department of Astronomy, University of California, Berkeley, CA 94720--3411, USA.}
\altaffiltext{6}{Department of Physics and Astronomy, University of Toledo, 2801 West Bancroft Street, Toledo, OH 43606, USA.}
\altaffiltext{7}{Computational Cosmology Center, Lawrence Berkeley National Laboratory, 1 Cyclotron Road, Berkeley, CA 94720, USA.}
\altaffiltext{8}{Institute for Astronomy, University of Hawai', 2680 Woodlawn Drive, Honolulu, HI 96822, USA.}
\altaffiltext{9}{Columbia Astrophysics Laboratory, Columbia University, New York, NY 10027, USA.}
\altaffiltext{10}{Einstein Fellow.}
\altaffiltext{11}{Spitzer Science Center, California Institute of Technology, Pasadena, CA, 91125, USA.}
\altaffiltext{12}{Dunlap Institute for Astronomy and Astrophysics, University of Toronto, 50 St. George Street, Toronto M5S 3H4, Ontario, Canada.}
\altaffiltext{13}{Caltech Optical Observatories, California Institute of Technology, Pasadena, CA 91125, USA.}
\altaffiltext{14}{University of California, Santa Barbara, CA 93106--9530}
\altaffiltext{15}{Visiting Astronomer at the Infrared Telescope Facility, which is operated by the University of Hawaii under Cooperative Agreement no. NNX-08AE38A with the National Aeronautics and Space Administration (NASA), Science Mission Directorate, Planetary Astronomy Program.}

\begin{abstract}
During a synoptic survey of the North American Nebula region,
the Palomar Transient Factory (PTF) detected an optical outburst (dubbed PTF10nvg)
associated with the previously unstudied flat or rising spectrum infrared source IRAS 20496+4354. 
The PTF $R$-band light curve reveals that PTF10nvg brightened by more than 5 mag during the current outburst, 
rising to a peak magnitude of $R_{\rm PTF} \approx 13.5$ in 2010 Sep.  Follow-up observations indicate PTF10nvg has undergone a similar $\sim$5 mag brightening in the K band, and possesses a rich emission-line spectrum, including numerous lines commonly assumed to trace mass accretion and outflows.  Many of these lines are blueshifted by $\sim$175 km s$^{-1}$ from the North American Nebula's rest velocity, suggesting that PTF10nvg is driving an outflow.   Optical spectra of PTF10nvg show several TiO/VO bandheads fully in emission, indicating the presence of an unusual amount of dense ($> 10^{10}$ cm$^{-3}$), warm (1500--4000 K) circumstellar material.  Near-infrared spectra of PTF10nvg appear quite similar to a spectrum of McNeil's Nebula/V1647 Ori, a young star which has undergone several brightenings in recent decades, and 06297+1021W, a Class I protostar with a similarly rich near--infrared emission line spectrum.  While further monitoring is required to fully understand this event, we conclude that the brightening of PTF10nvg is indicative of enhanced accretion and outflow in this Class-I-type protostellar object, similar to the behavior of V1647 Ori in 2004--2005.
\end{abstract}


\section{Introduction}

A defining characteristic \citep[e.g.,][]{Joy1945} of young stars is their
photometric variability.  At optical wavelengths, periodic low-amplitude 
variations are thought to arise from surface inhomogeneities on a rotating star; 
aperiodic large-amplitude variability is more often attributed to accretion-related activity.
The ``activity" amplitude generally declines as the star ages and
the disk evolves, with ``outburst" behavior exhibited by some of the most active sources. 
This has resulted in the definition of EXor type (actually named after EX Lup) young-star variables;
these objects often undergo repeated outbursts with 2--3 mag amplitudes and characteristic timescales of weeks to months \citep{Herbig2001,Herbig2008, Lorenzetti2009}.
More extreme are the FU Ori-type variables, which undergo 4--6 mag amplitude
increases over a few months, followed by a slow, decades-long decay \citep{Herbig1989, Hartmann1996}.  
In both cases, the variability is commonly attributed to nonsteady mass
accretion from a circumstellar disk.  Theoretical models typically invoke 
thermal, gravitational, or magnetorotational disk instabilities to explain
these accretion outbursts \citep{Bell1994, Kley1999,Armitage2001,Vorobyov2005,Boley2006,Zhu2009}, potentially 
triggered by binary interactions \citep{Reipurth2004b,Lodato2004}.
Recently documented outbursting young stars
include V1647 Ori \citep{McNeil2004,Briceno2004,Reipurth2004,Walter2004,Ojha2006,Fedele2007,Aspin2009}, V1118 Ori \citep{Lorenzetti2007,Audard2010}, V733 Cep \citep{Reipurth2007,Peneva2010}, ZCMa \citep{Szeifert2010,Whelan2010}, and EX Lup itself \citep{Herbig2001}.

\begin{figure*}
\epsscale{0.85}
\plottwo{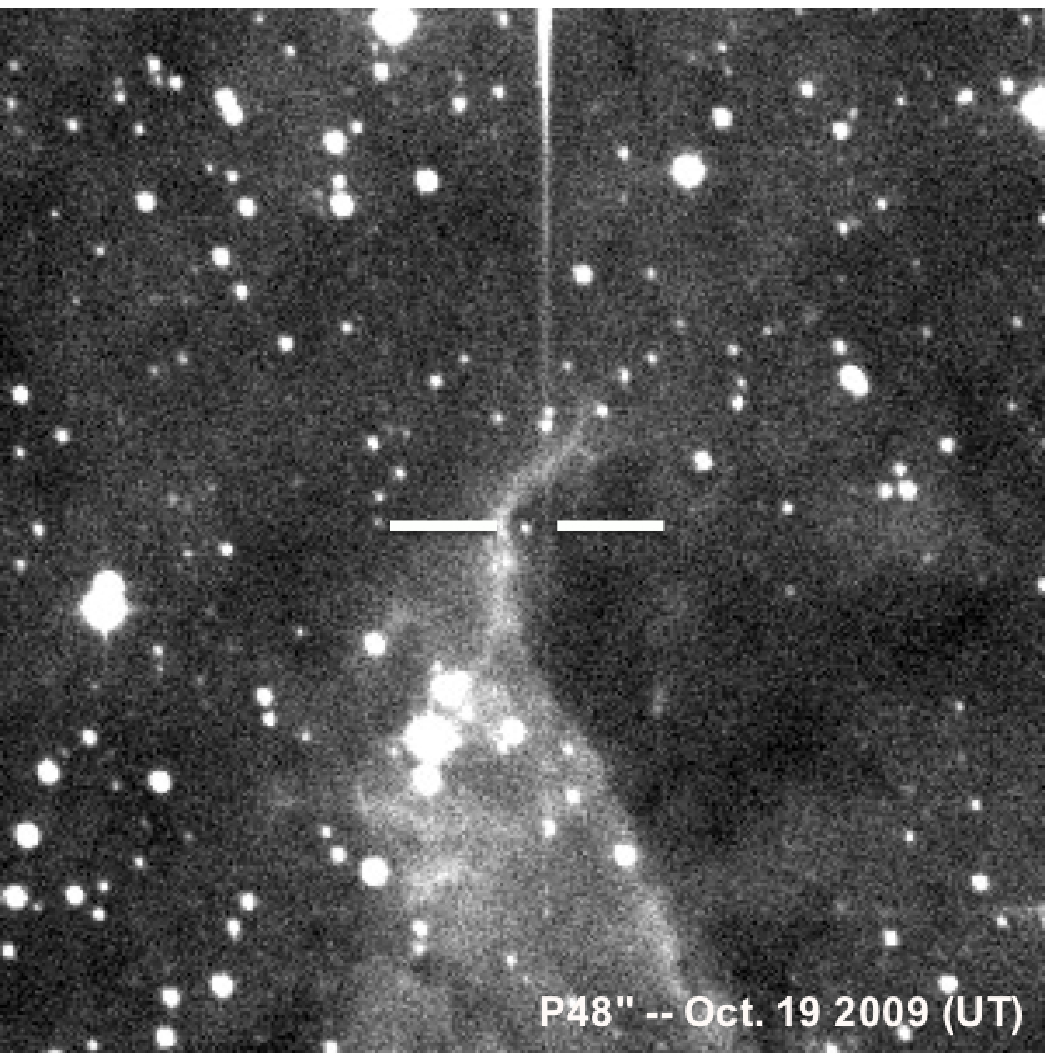}{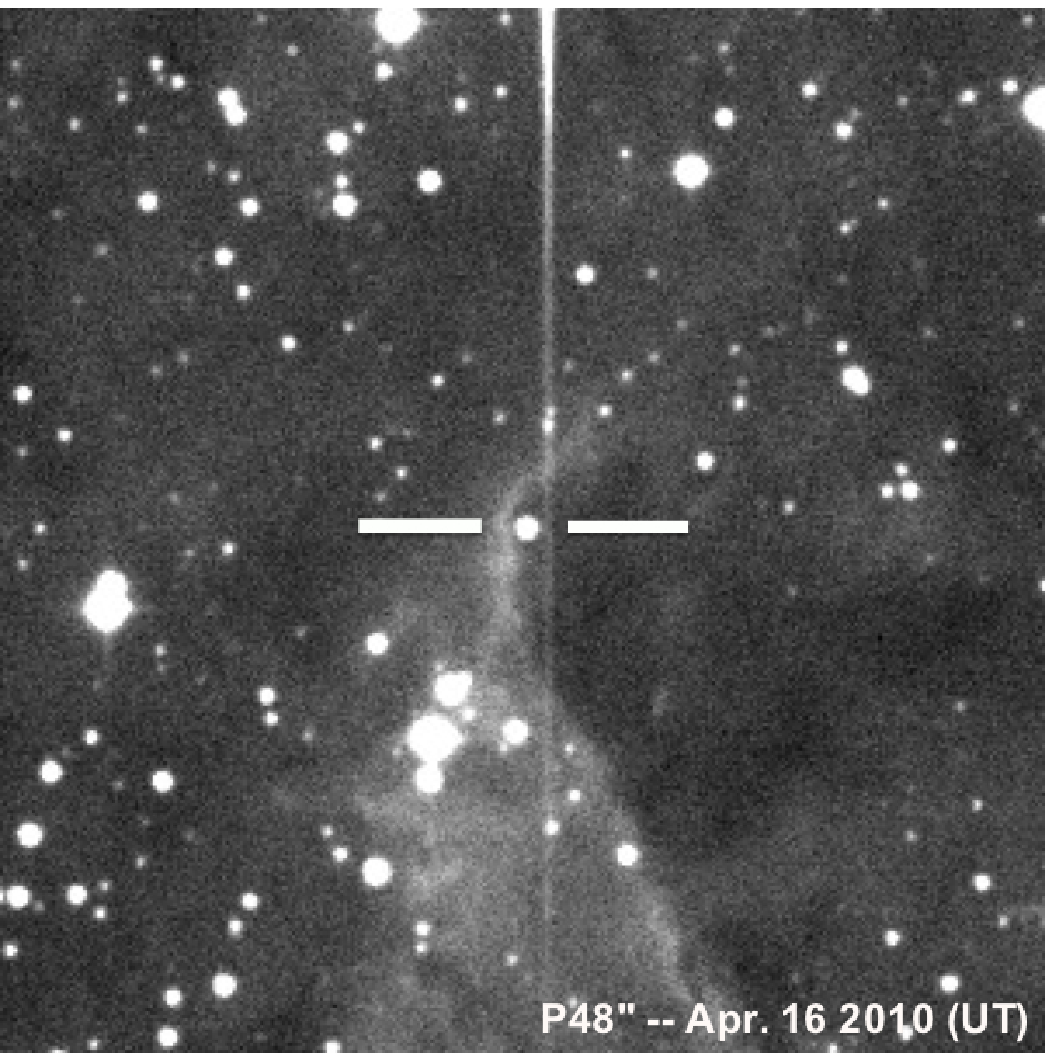}
\caption{ \normalsize{PTF images of PTF10nvg on 2009 Oct.\ 19 (left) and nearly 6 months later, on 2010 April 16 (right).  Images are 5$\arcmin$ on a side, with north up and east to the left; PTF10nvg is identified with white crosshairs.  The nebulosity in each $R$-band frame was pre-existing and is not associated with the outbursting behavior of the point source.  PTF10nvg is embedded within the molecular cloud associated with the North American/Pelican Nebula. A bright diffraction spike from a star just north of PTF10nvg does not substantially affect the PSF photometry derived for PTF10nvg. }}\label{fig:P48_frames}
\end{figure*}

With optical and near-infrared telescopes now regularly patrolling the sky, 
and ongoing variability surveys that target active star forming regions, 
the full gamut of young-star variability will soon be revealed, 
quantifying the characteristic time scales for eruptive events of varying amplitudes.
One such ongoing investigation is using the Palomar 48 inch Schmidt telescope and the Palomar Transient Factory
\citep[PTF;][]{Law2009,Rau2009} infrastructure to catalogue photometric variability
in the North American/Pelican Nebula region of recent star formation. 
Located at a distance of $\sim$600~pc, the population of young stars in this region 
(identified largely from surveys in H$\alpha$; e.g., 
\citealp{Herbig1958, Welin1973, Ogura2002}, or
the mid-IR, e.g., \citealp{Guieu2009}; Rebull et al., in prep.) is not as well studied as members of more proximate star-forming regions, but already includes a disproportionately large number of the known members
of the exclusive FU Ori class, namely V1057 Cyg and V1515 Cyg.
These two objects, along with FU Ori itself, are the defining members of the
class; they exhibit strong wind signatures in lines such as \iontoo{Na}{i} D and H$\alpha$,
metallic absorption patterns similar to those of FG-type supergiants in the optical
and M-type supergiants in the IR, and large thermal IR luminosity.

We have identified a new outbursting object located at position $\alpha = 20^{\rm h}51^{\rm m}26.23^{\rm s}$, $\delta = +44^\circ 05' 23.9''$
(J2000.0), just south of the Pelican Nebula.  It sits within the apex of a nebular arc
that wraps around the object from northwest to the east and then to the south.  The source is located within 1\arcsec\ of the position of IRAS 20496+4354, well inside the 16\arcsec\ error ellipse associated with that detection.  The source was further detected by the Midcourse Space Experiment (MSX) in 1996-1997, {\it Spitzer} in 2006 August 9--11 \citep[][Rebull et al., in prep.]{Guieu2009}, and AKARI/IRS during its 2006-2007 mission.  

At shorter wavelengths, the source was not visible ($K > 15.3$ mag) to the Two Micron All Sky Survey \citep[2MASS; ][]{Skrutskie2006}.  The 2-24 $\mu$m spectral index \citep[$\alpha$ = $d$ log($\lambda F_{\lambda}$) \ $d$ log $\lambda$][]{Lada1987} implied by the 2MASS $K$-band upper limit and the IRAS 24 $\mu$m detection is $\alpha \gtrsim$1.7, consistent with a designation as a heavily embedded Class I protostar in the classification system devised by \citet{Lada1987}.  The source does, however, appear in the $J$ and $K$ images of \citet{Magnier1999} that were obtained as follow-up observations
to the {\it IRAS} detection.   The source is not obvious on Digital Sky Survey (DSS) images (implying $m_{pg} > 21$ mag),
but NOMAD, using DSS-2 data, reports $m_B = 20.17$ and $m_R = 18.28$ mag for a source 3.4\arcsec\ away
from the source position reported here \citep{Zacharias2005}.  
Images of the field taken with H$\alpha$ and [S~II] filters by \citet{Bally2003} appear to show faint smudges at the source position.

We report here the outburst discovery data, as well as follow-up multi-filter
optical and near-infrared (NIR) photometric and spectroscopic monitoring that 
was obtained during various stages of the 2010 outburst.  We find evidence for
strong emission lines and broadly absorbed wind features, similar to other outbursting
young objects, but also some unique spectroscopic attributes, including molecular emission
from TiO and VO.  We analyze the data
in the context of an eruptive Class-I type, heavily embedded protostar. 

\section{Observations}\label{sec:Obs}

\subsection{Optical Photometry}\label{sec:PTF_phot}

During the 2009 and 2010 observing seasons, PTF obtained red optical images of the North American/Pelican Nebula star-forming region with a typical 5-day cadence.  These observations were conducted with the main PTF Survey Camera, the former CFHT12K mosaic camera now extensively re-engineered and mounted on the 48 inch Samuel Oschin telescope at Palomar Observatory (hereafter ``P48''). The camera is a mosaic of 11 CCDs, covering a 7.8 square degree field of view with 1\arcsec\ sampling; typical conditions at Palomar Observatory produce 2.0\arcsec\ full width at half-maximum intensity (FWHM) images \citep{Law2009}.  The $R_{\rm PTF}$ filter, a Mould $R$ band, is similar to SDSS-$r$ in shape but is shifted redward by $\sim 27$~\AA\ and is $\sim 20$~\AA\ wider.  The typical 5$\sigma$ limiting magnitudes are $m_{R} \approx 20.6$ (AB) in 60~s exposures.  Representative images of PTF10nvg from the 2009 and 2010 observing seasons are shown in Figure \ref{fig:P48_frames}.

\begin{figure*}
\epsscale{1.0}
\plotone{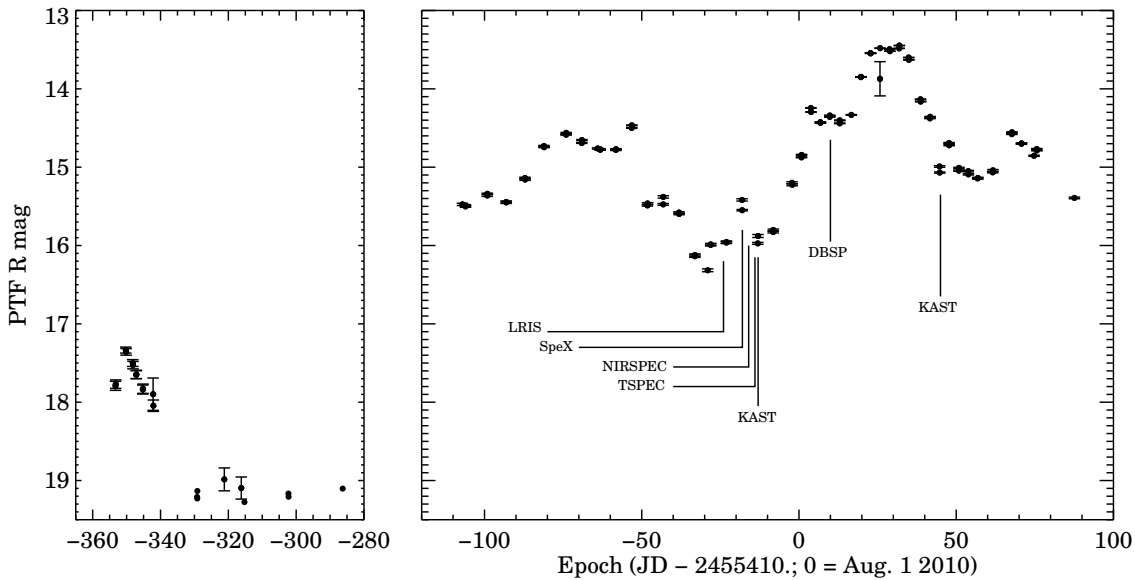}
\caption{ \normalsize{R$_{PTF}$-band light curve of PTF10nvg, measured via PSF photometry of individual P48 frames (see \S \ref{sec:PTF_phot} for details).  Left panel shows observations from the 2009 season; right panel shows 2010 observations, with labels identifying the timing of the spectra presented in \S \ref{sec:opt_spec} and \ref{sec:nir_spec} \label{fig:P48_LC} } }
\end{figure*}

Transient sources are detected in PTF monitoring data by means of automated reduction pipelines, including a near-real-time image-subtraction pipeline hosted at Lawrence Berkeley National Laboratory (LBNL).  Well-detected sources in the difference images are scored (using a human trained machine-based classifier) for their likelihood of being truly astrophysical in nature or of spurious origin.  
Variable sources with larger likelihoods of being nonspurious are passed to an automatic source classifier at UC Berkeley (``Oarical"), which combines PTF measurements with all other available information (e.g., SIMBAD identifications, 2MASS photometry, etc.) to provide probabilistic classifications of PTF detections (Bloom et al. 2010, in prep).  These initial classifications are made available to PTF collaboration members via the PTF Follow-up Marshal, which enables visual inspection of current and reference images, precursor PTF light curves, and any subsequent PTF spectroscopy. 

The source reported here was detected by the PTF pipeline and automatically assigned the name 10nvg, following PTF naming conventions.  PTF images are typically re-reduced for sources of particular interest, such as PTF10nvg, via a ``white-glove pipeline,'' where PSF photometry is performed on individual frames.  The PSF photometry is calculated using a modified version of the pipeline developed by the Supernova Legacy Survey \citep[see][]{Astier2006}.  Absolute calibration is performed relative to the USNO-B1 catalog \citep{Monet2003}, and typically has an uncertainty of $\sim 0.15$ mag.  The full P48 light curve of PTF10nvg produced by this ``white-glove reduction'' is shown in Figure \ref{fig:P48_LC} and the photometry appears in Table \ref{tab:P48}.

\begin{figure}
\epsscale{0.8}
\plotone{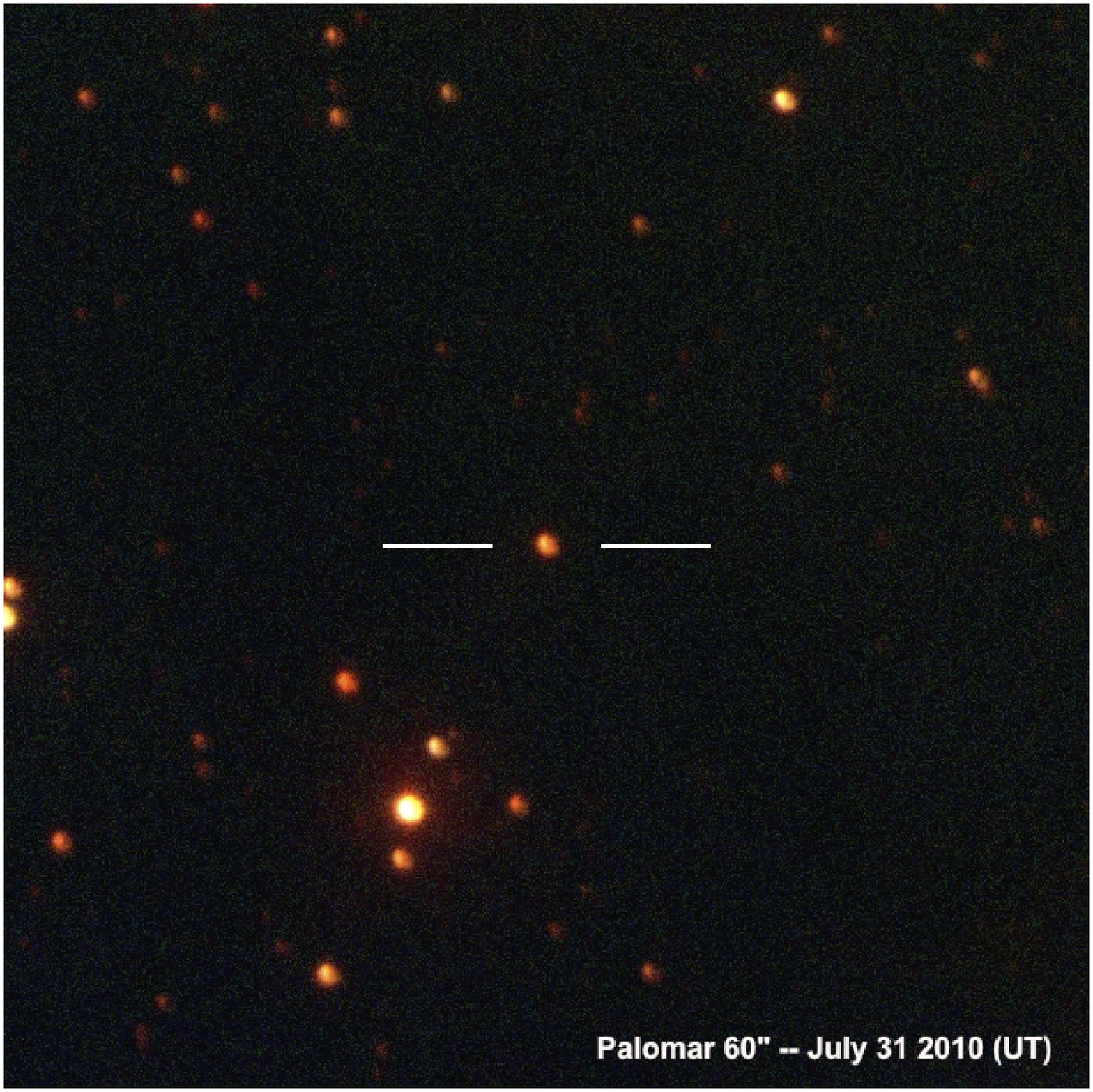}
\caption{ \normalsize{Three-color frame of PTF10nvg $riz$ images obtained with the P60. The frame is 4.1\arcmin\ on a side, with north up and east to the left. }}\label{fig:P60_frame}
\end{figure}

The PTF also makes use of the robotic Palomar 60 inch telescope \citep[P60;][]{Cenko2006} to obtain multi-color photometry for source verification and classification purposes.  PTF10nvg was observed by the robotic Palomar 60-inch starting 2010 August 1.35 in the $r$ band, $i$ band and $z$ band filters. The absolute zeropoint calibration was derived relative to the Sloan Digital Sky Survey
from separate fields observed by the Palomar 60-inch during the
night of 2010 Sep 3 with the same filter. All other fields are
calibrated relative to this night. Figure \ref{fig:P60_frame} presents a three-color image constructed from the P60 $riz$ frames, and Table ~\ref{tab:P60} contains the individual measurements.

We note that there is a $\sim$0.45 mag offset between the P48 $R_{\rm PTF}$ and P60 $r$ band light curves for PTF10nvg.  A simple zero-point error may contribute to this offset, but it is likely dominated by differences in each telescope's filter+detector response, and the different photometric systems underlying their calibration (Vega $R$ vs. AB magnitude $r$).  Given the multiple possible contributions to this offset, we have chosen to preserve each light curve on its native system, rather than explicitly transform them onto a common system. 

\subsection{Near-Infrared Photometry}\label{sec:NIR_phot}

\begin{figure*}
\epsscale{0.8}
\plottwo{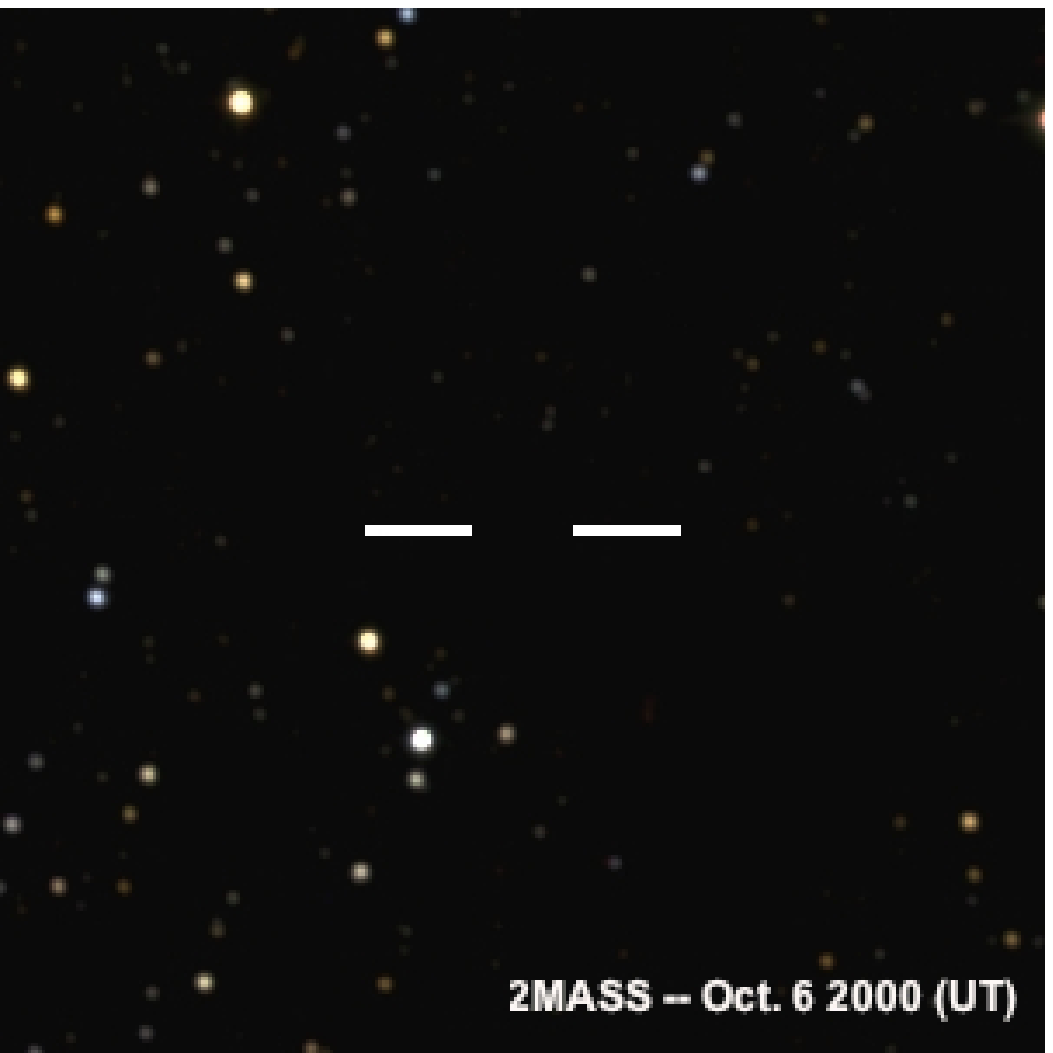}{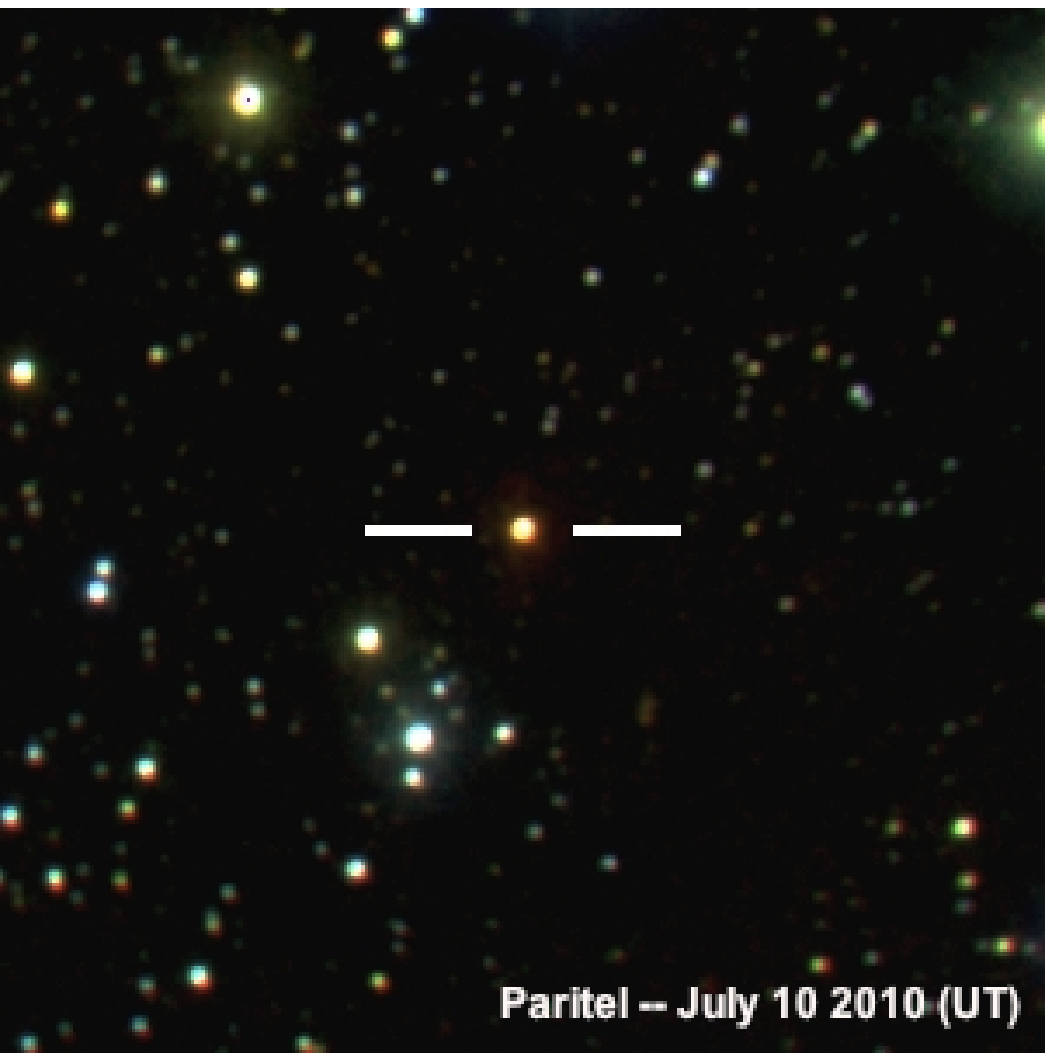}
\caption{ \normalsize{Three-color JHK images of the field containing PTF10nvg obtained by 2MASS (left) and PAIRITEL (right).  Frames are 5$\arcmin$ on a side, with North up and East to the left.  PTF10nvg is not detected in the 2MASS imaging acquired on 2000 Oct. 6, but is visible in the PAIRITEL imaging acquired 2010 July 10.}}\label{fig:2MASS_Paritel_frames}
\end{figure*}

Near-infrared observations of PTF~10nvg were conducted with the 1.3~m Peters Automated Infrared Imaging Telescope \citep[PAIRITEL; ][]{Bloom2006} starting on 2010 July 10. PAIRITEL is a roboticized system using the former 2MASS southern hemisphere survey camera that employs two dichroics to observe simultaneously in the $J$, $H$, and $K_s$ bands. Observations were scheduled and executed via a robotic system. PAIRITEL is operated in a fixed observing mode in which 7.8 s double-correlated ÒimagesÓ are created from the difference of a 7.851 s and a 51 ms integration taken in rapid succession \citep[see ][]{Blake2008}. The standard observing procedure involves taking three image pairs prior to dithering the telescope.

The raw data from these images are reduced using standard IR reduction methods via \hbox{PAIRITEL} PIPELINE III and the flux for all sources is measured via aperture photometry using SExtractor \citep{Bertin1996}, calibrated against 2MASS.  PTF10nvg saturates the 7.851 s frames; however, PIPELINE III produces ``short-frame'' mosaics consisting of reduced, stacked 51 ms images \citep[see also][]{Bloom2009}. The ``short-frame'' mosaics contain $>$10 bright 2MASS stars which we use to properly calibrate photometric measurements of PTF 10nvg in these images. PAIRITEL has a known systematic uncertainty of $\sim$0.02--0.03 mag in each of the  $J$, $H$, and $K_s$ bands \citep[see also][]{Blake2008,Perley2010}, which, in the case of PTF10nvg, is larger than the statistical error in all three bands. Thus, we add a systematic error of 0.03 mag in quadrature with the systematic uncertainty to determine the total uncertainty in each band. The $JHK_s$ images of PTF10nvg obtained by PAIRITEL on 2010 July 10 are shown as a three-color frame in Figure \ref{fig:2MASS_Paritel_frames}; all $JHK_s$ magnitudes measured for PTF10nvg by PAIRITEL are presented in Table \ref{tab:Paritelmags}.  

We also examined the 2MASS data products to determine if PTF10nvg was detected during that survey. A query of the 2MASS source catalog via the GATOR interface does not identify a source within 10\arcsec\ of this position, and visual inspection of the 2MASS images confirm that the source was undetected in 2000 October, when the field was imaged. Analysis of the 2MASS frames provides 5$\sigma$ upper limits on the brightness of the source at the epoch of the 2MASS observations of $J>$16.35, $H>$15.38, and $K_s>$14.80; we include these upper limits in the $JHK_s$ photometry presented in Table \ref{tab:Paritelmags}.  

\subsection{Optical Spectroscopy}\label{sec:opt_spec}

Moderate-resolution optical spectra of PTF10nvg were obtained on four occasions: 2010 July 8 with the Low Resolution Imaging Spectrometer \citep[LRIS;][]{Oke1995,McCarthy1998,Steidel2004} on the Keck-I 10~m telescope, 2010 July 19 and 2010 Sep. 16 with the Kast double spectrograph \citep{Miller1993} on the Shane 3~m telescope at Lick Observatory, and 2010 Aug. 11 with the Double Spectrograph \citep[DBSP;][]{Oke1982} on the Hale 5~m telescope at Palomar Observatory.  

LRIS observations were made by J.S.B., A. Cucchiara, A.N. Morgan and D.A. Perley with the D560 dichroic, which has a 50\% transmission split at $\lambda \approx 5700$~\AA, and a 1.0\arcsec\ long slit. The blue side of the spectrograph was configured with the 600/4000 grism, providing $\sim$4~\AA\ resolution over the 3010--5600~\AA\ wavelength range.  The red spectrograph utilized the 400/8500 grating, providing $\sim$7~\AA\ resolution over 5600--10300~\AA.  The LRIS spectra were reduced in the IRAF\footnote{IRAF is distributed by the National Optical Astronomy Observatory, which is operated by the Association for Research in Astronomy, Inc., under cooperative agreement with the National Science Foundation.} environment using standard routines. Cosmic rays were removed using the LA Cosmic routine \citep{v01}. Spectra were extracted optimally \citep{h86} and wavelength calibration was performed first relative to arc lamps and then tweaked based on night-sky lines in each individual image. Both air-to-vacuum and heliocentric corrections were then applied to all spectra. Extracted spectra were divided through by a smoothed flux standard to remove narrow-band ($< 50$~\AA) instrumental effects \citep{b99}. Finally, telluric absorption features were removed using the continuum from spectrophotometric standards \citep{wh88,mfh+00}.  

The July 19 Kast observations were obtained by C.V. Griffith and M.T. Kandrashoff, with reductions performed by J.M.S.; the Sept. 16 observations were obtained by S.B.C. and M.T. Kandrashoff, with reductions performed by S.B.C.  Kast was configured with a 2\arcsec\ slit, a
600/4310 grism on the blue side, and a 300/7500 grating on the red
side, yielding resolutions of $\sim 4$ and
$\sim 10$~\AA, respectively.  

DBSP data were acquired by L.M.R. using the D55 dichroic to obtain spectra with a 600~line mm$^{-1}$ grating blazed at 4000~\AA\ from the atmospheric cutoff to 5500~\AA\ at 1.1~\AA\ resolution, and with a 158 line mm$^{-1}$ grating blazed at 7500~\AA\ over the range 6300--8800~\AA\ at 2.5~\AA\ dispersion.  DBSP data reduction was performed by L.A.H., including flat fielding, extraction, and wavlength calibration, using standard routines in the IRAF environment.

\subsection{Near-Infrared Spectroscopy}\label{sec:nir_spec}

Moderate-resolution NIR spectra of PTF10nvg were obtained on 2010 July 14 with the SpeX spectrograph \citep{Rayner2003} at the NASA Infrared Telescope Facility, and on 2010 July 18 with the TripleSpec Spectrograph \citep{Wilson2004} at the Apache Point Observatory.  

SpeX observations were obtained by J.R. on 2010 July 14 with a 0.3\arcsec\ slit under clear skies and good seeing (0.4--0.5\arcsec). Both the SXD and LXD modes were utilized, providing nearly contiguous coverage over the range 0.8--2.5 $\mu$m with a spectral resolution of $R = 2000$, and over 3--4.2 $\mu$m ($L$ band) and 4.5--5.0 $\mu$m ($M$ band) at $R\approx 2500$.  PTF10nvg was nodded along the slit during the observation, such that differencing consecutive images effectively removes the sky emission.  The spectra were differenced, flat fielded, extracted, and wavelength calibrated using the SpeXTool IDL routines \citep{Cushing2004}.  Telluric correction and absolute flux calibration were performed using a spectrum of a nearby A0V star obtained at similar airmass and time to PTF10nvg and reduced in similar fashion; they were applied using the XTELLCOR IDL package \citep{Vacca2003}.

TripleSpec observations were obtained by K.R.C. on 2010 July 18 with the 0.7\arcsec\ slit, providing nearly contiguous coverage over 1.0--2.5 $\mu$m at $R \approx 5000$.   Spectra were obtained and reduced in the same manner as the SpeX observations, using a version of the SpeXTool package modified for use with APO/TripleSpec data.  

A NIRSPEC spectrum covering 0.95--1.12 $\mu$m at $R \approx 25,000$ was obtained on 2010 July 16 by L.A.H.
Standard data-reduction steps including wavelength calibration and image rectification, extraction of one-dimensional spectra,
and removal of telluric emission and absorption features was carried out by W.J.F. within the REDSPEC IDL package
developed by S. S. Kim, L. Prato, and I. McLean. 

\section{Discovery and Characterization}

PTF10nvg was first identified as an optical transient by the PTF automatic discovery and classification codes on 2010 July 8.   It was selected for spectroscopic follow-up observation as a bright event observable with Keck during morning twilight of July 8, and subsequently visually identified as a likely outbursting young star.  Analysis of the PTF data and the follow-up photometry and spectroscopy is presented below.

\subsection{Photometric Analysis}

\subsubsection{Outburst Characterization from the Optical/PTF Light Curve}

Although PTF10nvg is not known historically as an optical source, it was detected
by PTF at $m_R \approx 18$ mag in mid-2009 but then faded to $\sim 19.25$ mag on a time scale
of roughly 2 months.  It brightened again by more than 4 mag in less than 200 days,
while the field was too close to the Sun for observations.  PTF10nvg reached $m_R = 15$ early in 2010, faded to 16.3 mag within 40 days of the first peak, and then brightened again to 13.5 mag before fading to $\sim$15.5 mag in early Nov. 2010.  The two brightness peaks, as well as the intermediate lull, are all $\sim 50$ days in width.

\subsubsection{Pre- and Post-Outburst SED}

\begin{figure}
\epsscale{1.0}
\plotone{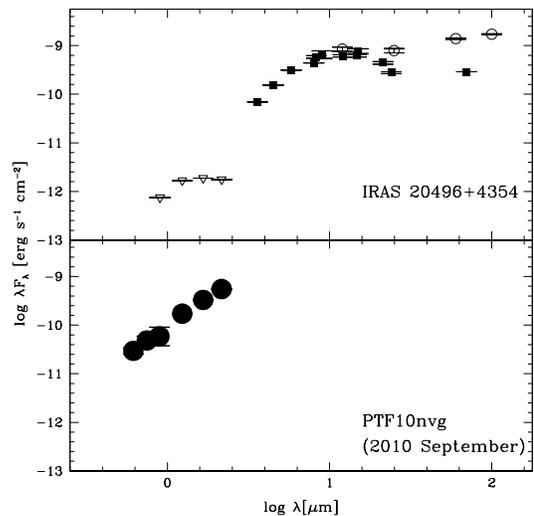}
\caption{ \normalsize{Pre-outburst (top panel) and post-outburst (bottom panel)
SED of PTF10nvg. The pre-outburst data derive from
KPNO 0.9~m optical and 2MASS NIR upper limits (triangles), 
{\it Spitzer} IRAC (3.6, 4.5, 5.8, 8.0 $\mu$m; filled squares) and MIPS (24, 70 $\mu$m; filled squares),
MSX (8.3, 12.1, 14.6, 21.3 $\mu$m; filled squares), AKARI (9 and 18 $\mu$m; filled squares), 
and {\it IRAS} (12, 25, 60, 100 $\mu$m; open circles) photometric measurements (note that the {\it IRAS} beam
was larger than that of the other instruments).  The post-outburst data
are those of 2010 July 9 as reported in Tables 1 and 3, when the object was in
a relative lull between its two optical brightness maxima (see Fig. \ref{fig:P48_LC}).}}\label{fig:SED}
\end{figure}

Pre-outburst photometry that we have collated from the literature is 
presented in Table \ref{tab:precursor_phot}, and the SEDs assembled 
from this pre-outburst photometry and our own post-outburst detection 
data are displayed in Figure~\ref{fig:SED}.  

From 3 to 10 $\mu$m the SED is rising, then becomes roughly flat from 
10 to 100 $\mu$m.  The integrated luminosity of the pre-outburst IR source 
between 3 and 100 $\mu$m is 25~L$_\odot$; extrapolating the SED longward of the IRAS 
100 $\mu$m measurement as a standard Rayleigh-Jeans function, and integrating to $\lambda=\infty$, 
increases the luminosity by $\sim$40\%.  Other Class I sources, however, have SEDs that peak at or slightly longward of 100 $\mu$m \citep[e.g. ][]{Enoch2009}; if PTF10nvg's SED possesses a similar peak, the total luminosity would be larger than this, which assumes a sharply declining power-law directly from 100$\mu$m.  
The integrated luminosity of the outbursting source between 0.7 and 2.2 $\mu$m 
is 2.5~L$_\odot$, compared to $<$0.016 L$_\odot$ before the outburst.
There is no post-outburst photometry longward of 2.2 $\mu$m.
 
 \begin{figure*}
\epsscale{1.0}
\plottwo{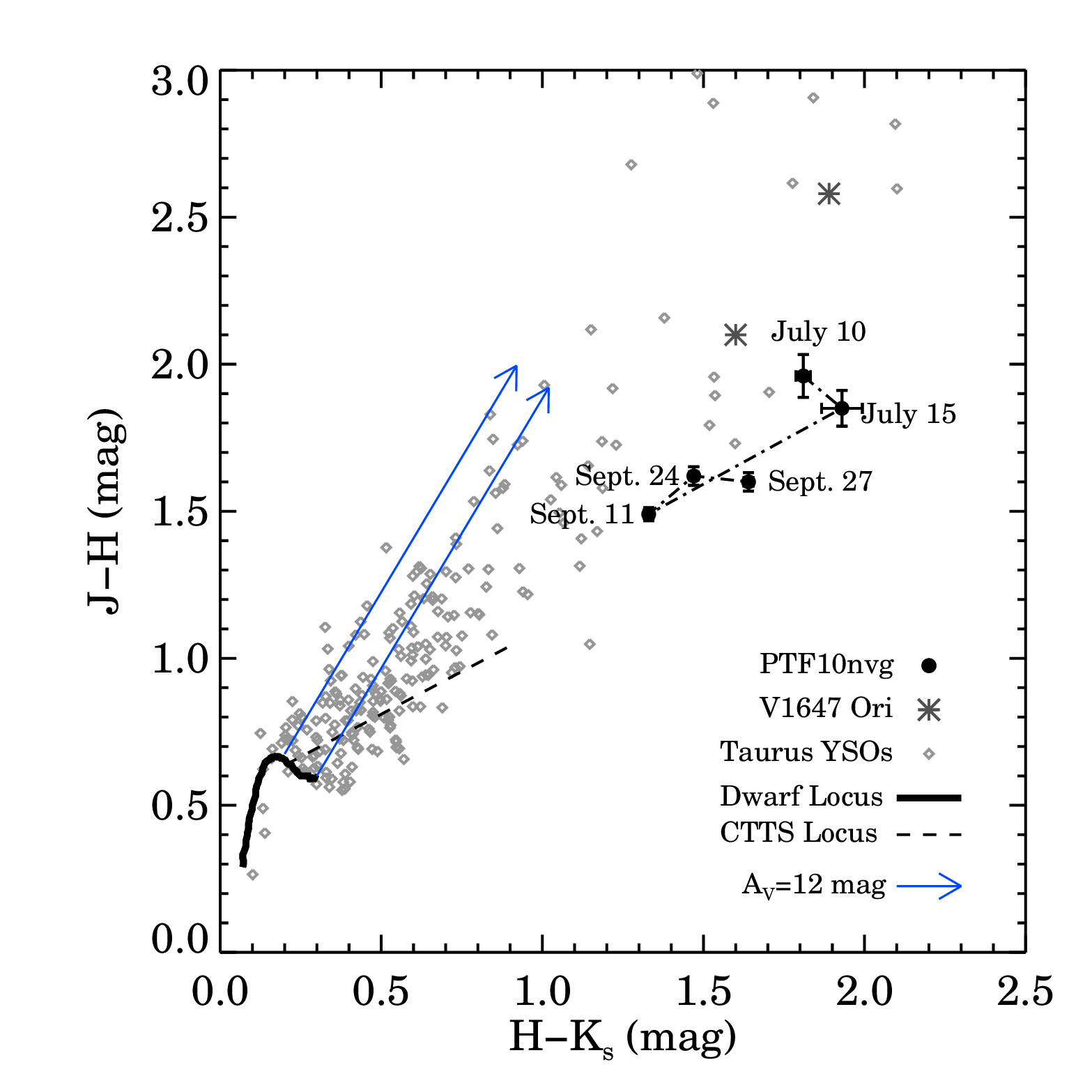}{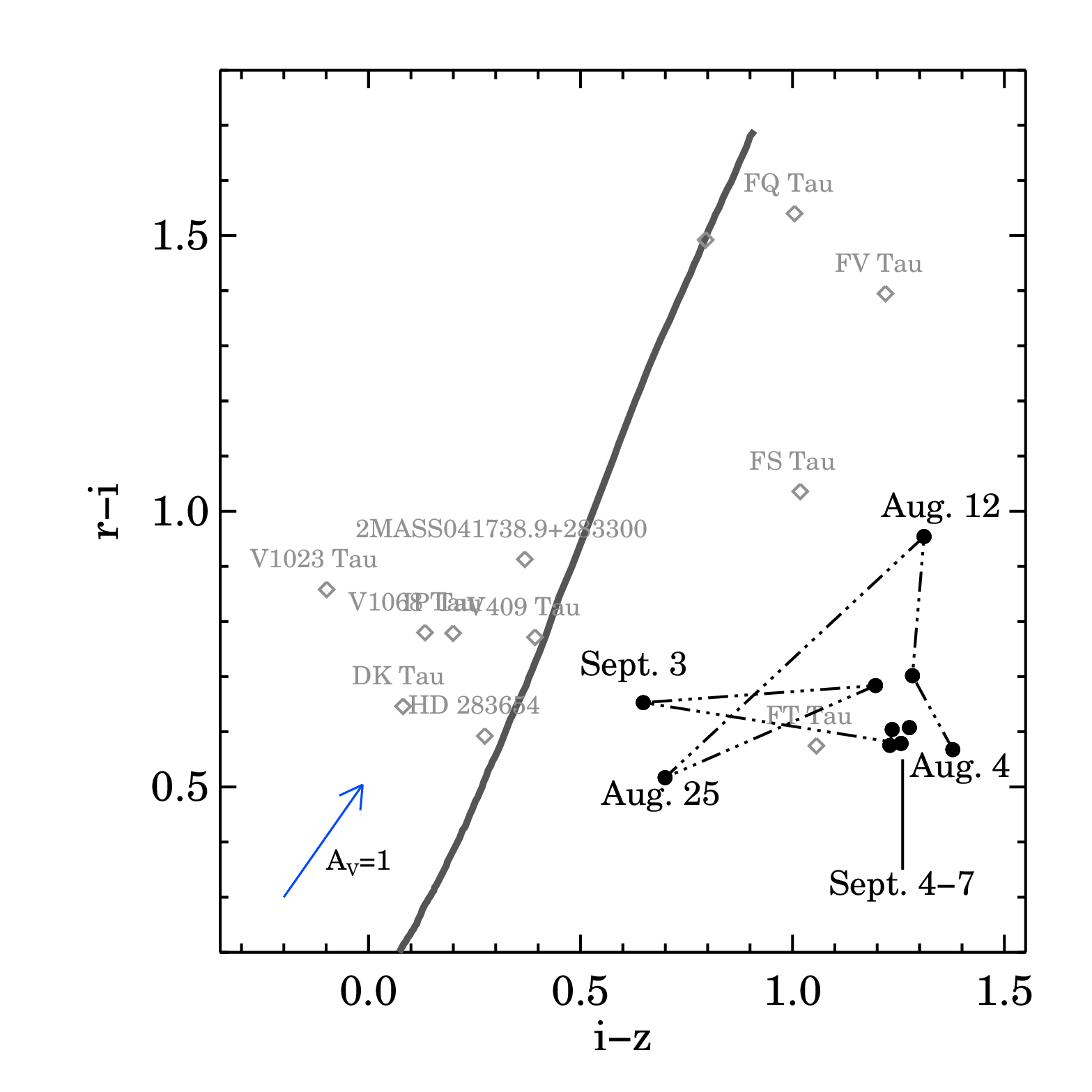}
\caption{ \normalsize{Location of PTF10nvg in $JHK_s$ (left) and $riz$ (right) color-color diagrams.  Solid lines in each panel indicate the locus of SDSS/2MASS colors for main-sequence field dwarfs \citep[F to M spectral types; ][]{Covey2007}; the dashed line in the top panel shows the Classical T Tauri locus derived by \citet{Meyer1997} and transformed into the 2MASS photometric system using the relationship derived by \citet{Carpenter2001}.  Class I, II, and III objects in Taurus, as well as the outbursting object V1647 Ori, are shown for comparison; photometry for these comparison stars was catalogued by \citet{Finkbeiner2004}, \citet{Reipurth2004}, and \citet{Rebull2010}.}}\label{fig:colors}
\end{figure*}
 
The optical and NIR colors of PTF10nvg, shown in Figure~\ref{fig:colors}, are extremely red during outburst: $r-i \approx 0.6$ mag, $i-z \approx 1.1$ mag, $J-H \approx 1.9$ mag, and $H-K_s \approx 1.9$ mag.  These colors are within the distribution but toward the blue end of NIR colors exhibited by Class-I-type stars in Taurus, and significantly redder than can be explained by a range of disk accretion, inner hole, and inclination properties along the Class II ``Classical T Tauri locus" \citep{Meyer1997}.  In 2010 July, PTF10nvg appeared redder in $H-K_s$ and bluer in $J-H$ than most known ``FU Ori-like" objects \citep[e.g., see Figure 4 by][]{Greene2008}.  
As PTF10nvg began fading in Sept. 2010, however, its observed $JHK_s$ colors became somewhat bluer, comparable to the colors of V346 Nor, a previously known ``FU Ori-like'' object.  The $JHK_s$ colors of PTF10nvg are also similar to those of V1647 in Feb--March 2004, early in the 2004--2005 outburst.

High energy radiation has been detected from other outbursting young stars 
\citep[e.g., V1647;][]{Kastner2006}, but no X-ray detection can be confidently associated with PTF10nvg 
at this time.  X-ray observations of PTF 10nvg were obtained with the {\it Swift} X-Ray Telescope \citep[XRT;][]{Burrows2005} starting on 2010 Aug.\ 29 for a total exposure time of 3.6 ks. We extract the 0.5--8.0 keV counts from an extraction region of 64 pixels, $\sim$2.5 arcmin, where we fit the point spread function model \citep[see][]{Butler2007} at the centroid of the optical position of PTF 10nvg. We detect no X-ray emission from PTF 10nvg. Assuming a $\Gamma = 2$ power law spectrum and a column density of $N_{\rm H}$ = $10^{22}$ cm$^{-2}$ we find a 3-$\sigma$ upper limit of the flux: $F_X < 1.2 \times 10^{-13}$ erg cm$^{-2}$ s$^{-1}$.  Consistent with both the sensitivity of the  ROSAT All Sky Survey \citep{Voges1999} and the Swift upper limit, the closest ROSAT source is approximately 
0.9\arcmin\ SW of PTF10nvg's optical position, suggesting that it is 
unlikely to be a counterpart to the star.

\subsection{Spectroscopic Analysis}

PTF10nvg exhibits a rich set of emission-lines, including hydrogen and both low- and high-excitation metallic atomic lines.  Atomic line-strength differences between the multiple optical and IR observations are presented quantitatively in Tables \ref{tab:opt_lines} and \ref{tab:nir_lines}.  Equivalent widths and line fluxes reported in these tables are calculated by integrating across the line after subtracting off the local continuum, which is estimated by interpolating between two line free regions well separated from the feature in question (typically v $>$ 1000 km s$^-1$ away).  Line fluxes are simply calculated as the direct integral of the line in the flux calibrated spectra; equivalent widths are calculated by dividing the integrated flux by the mean value of the local continuum.   

\begin{figure*}
\epsscale{1.0}
\plotone{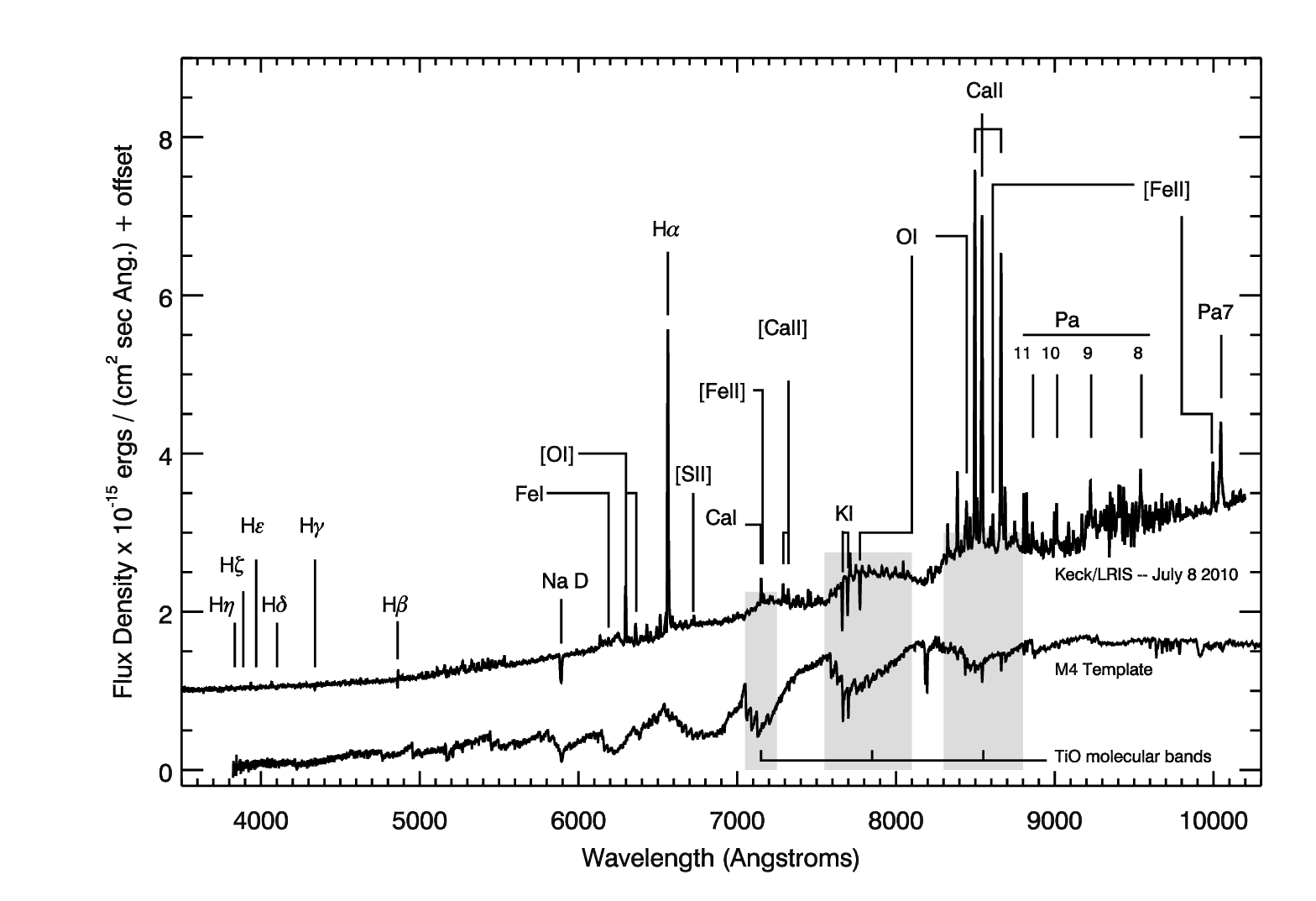}
\caption{ \normalsize{Optical spectrum of PTF10nvg obtained with Keck/LRIS on 2010 July 8.  Strong emission lines are identified, with line strengths presented quantitatively in  Table \ref{tab:opt_lines}.  An M4 field-dwarf spectral template \citep{Bochanski2007} is shown to illustrate the correspondence between the TiO emission bands in PTF10nvg and typical TiO absorption bands in cool M-type photospheres.
}}\label{fig:PTF10nvg_OptSpec}
\end{figure*}

\begin{figure*}
\epsscale{1.2}
\plotone{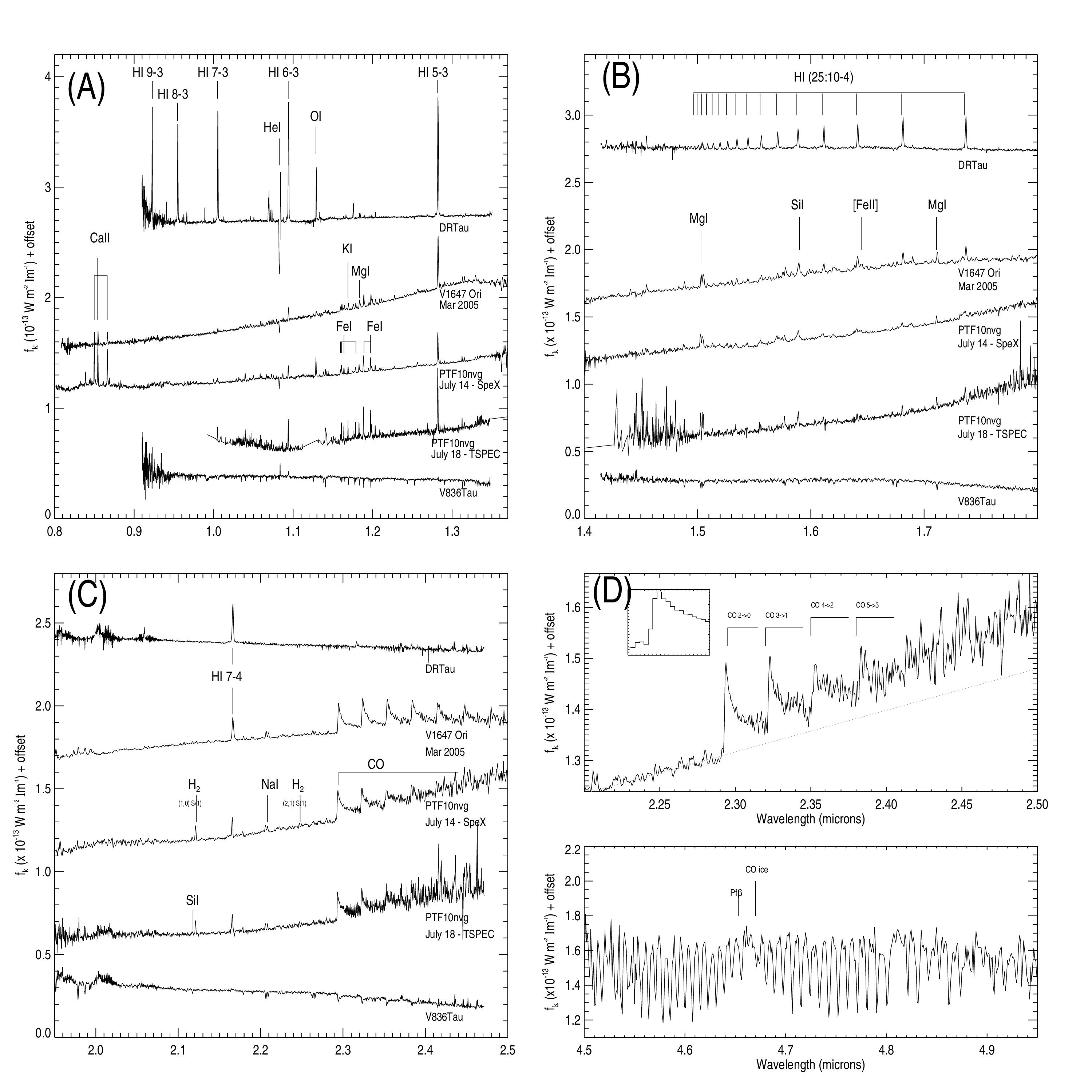}
\caption{ \normalsize{$YJ$-, $H$-, $K$-, and $M$-band spectra of PTF10nvg, in panels A, B, C, and D, respectively.  
Panels A-C also include spectra of DR Tau (heavily accreting Class II object), V836 Tau (typical Class III object), and V1647 Ori (McNeil's Nebula outburst) for comparison.  Strong emission lines are labelled where possible; PTF10nvg line strengths are also presented quantitatively in Table \ref{tab:nir_lines}.  Notably, the equivalent widths of the emission lines in PTF10nvg are 50--100\% larger in the APO TripleSpec observation than in the SpeX observation even though the line fluxes are similar.  The difference in equivalent widths therefore likely reflects differences in the continuum flux level between the two epochs.}}\label{fig:PTF10nvg_IRSpec}
\end{figure*}

In addition to the many atomic features in emission in PTF10nvg's spectrum, there is also strong molecular emission from TiO and VO at red optical wavelengths (see PTF10nvg's discovery spectrum, shown in Figure~\ref{fig:PTF10nvg_OptSpec}), and from VO, H$_2$O, and CO in the IR regime (see PTF10nvg's SpeX and TripleSpec spectra, shown in Figure~\ref{fig:PTF10nvg_IRSpec}). To the best of our knowledge, this is the first time such prominent molecular emission has been detected from a young star at optical wavelengths.   

\begin{figure}
\epsscale{1.23}
\plotone{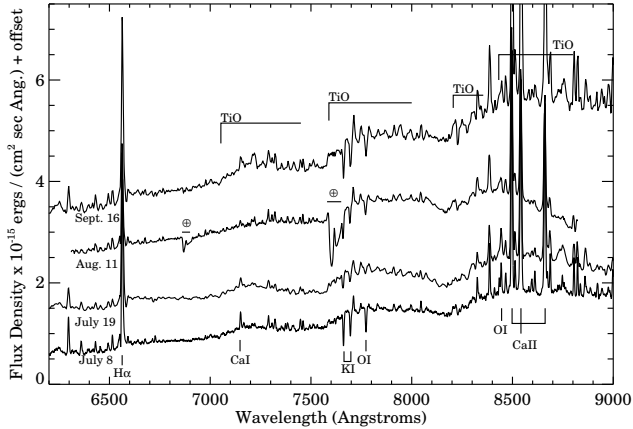}
\caption{ \normalsize{Comparison of the four optical spectra between the [O~I] and the Ca~II triplet region (note that the 11 August Palomar DBSP spectrum has not been corrected for telluric absorption.  The apparent ``continuum" differences are due to variations in the strength of the TiO emission over the four epochs. }}\label{fig:OptSpec_compare}
\end{figure}

The four optical spectra from Keck, Lick, and Palomar are compared over the [O~I] to Ca~II triplet region in Figure~\ref{fig:OptSpec_compare}.  Note that the spectra were obtained after the first brightness peak in the 2010 light curve, near the beginning and the end of the $m_R \approx 16.5$ mag nadir (Keck and Lick) and then when the object was brightening again and had reached $m_R \approx 14.8$ mag for the second time during our monitoring.   There is some evidence for a change in the overall continuum slope during this month; as the object brightened by several magnitudes it also became redder in a manner similar to interstellar extinction by $\sim$1 mag in the blue and $\sim$1.5 mag in the red.  The emission feature strengths also varied, as discussed below.  The broad TiO band emission, also discussed below, appeared strongest in the second (Lick) spectrum just before the object began its second brightening.

We analyze spectroscopic features of particular interest in more detail below.  To analyze the kinematic structure of the observed line profiles, we assume a rest velocity for PTF10nvg of $V_{LSR} \sim 4$ km s$^{-1}$, consistent with $^{13}$CO observations of the surrounding molecular gas \citep{Dobashi1994}.  

\subsubsection{H~I Lines}\label{sec:hi}

PTF10nvg has strong H~I emission lines including H$\beta$, H$\alpha$, the Paschen series (detectable up to 21), and the Brackett series including Br-$\alpha$ at 4.05 $\mu$m.  No Balmer jump is apparent, but the pileup of the higher-order Brackett lines is visible in the $H$ band of the SpeX observations. The high-dispersion NIRSPEC spectrum reveals profile widths in the  Pa-$\gamma$ and  Pa-$\delta$ H~I lines of $\sim$125 km s$^{-1}$ FWHM.  The lines are slightly asymmetric with respect to their peak emission at roughly zero velocity, having somewhat more integrated flux on the blue side of line center than on the red side.   The H$\alpha$ line equivalent widths (EWs) are $-$26 to $-$46~\AA\ among the four optical spectra, moderate but not extreme relative to Taurus Class I or Class II objects \citep{White2004}. 

We have derived an estimate of the extinction towards PTF10nvg from the strength of its Pa $\delta$ and Br $\gamma$ emission lines.  These lines, representing the 7-3 and 7-4 transitions respectively, share the same upper level.  Neglecting collisional effects (a questionable assumption that we will return to below), the relative frequency of radiative de-excitations from the same upper level should be determined entirely by atomic physics.  That is, the intrinsic flux ratio of these lines can be calculated in the collisionless limit as

\begin{equation}
\frac{F_{Br \gamma}}{F_{Pa \delta}} = \frac{A_{Br \gamma} \lambda_{Pa \delta}}{A_{Pa \delta} \lambda_{Br \gamma}}
\end{equation}

\noindent where F, A, and $\lambda$ give the flux, Einstein coefficient, and wavelength of each line.  Using line constants from the NIST atomic database, we calculate an intrinsic Br $\gamma$-to-Pa $\delta$ line ratio of 0.42 for a collisionless emission region.  Densities in magnetospheric accretion columns, however, are sufficiently high that collisions are likely to be an important factor for the emergent fluxes of HI emission lines.  Detailed radiative transfer models of relatively cool (T$_{max} =$ 6000 K) magnetospheric accretion columns nonetheless return a Br $\gamma$-to-Pa $\delta$ line ratio very close to this naive calculation.  Models of significantly warmer (T$_{max} = $9000 K) accretion columns return a larger line ratio of 2.21 (R. Kurosawa, priv. communication).  The line strength ratio in the SpeX spectrum of PTF10nvg is $\sim$8.2, demonstrating the presence of significant extinction relative to either accretion models.  We therefore estimate the extinction to the H~I line emission region (which may or may not be coincident with the stellar photosphere) by assuming a standard $R_V=$3.1 extinction law, and de-reddening the observed line fluxes to match each of these predicted intrinsic line ratios. These estimates place crude bounds on the extinction to PTF10nvg's emission line region of 6 $< A_V <$ 12.4.

Methods have also been developed to estimate a young star's mass accretion rate from the strength of its NIR H~I emission lines \citep[e.g., ][]{Muzerolle1998}.  These relationships, which link observed H~I line luminosities with mass accretion rates inferred from UV continuum excesses, have only been calibrated for observations of \hbox{T Tauri} stars; applying these relationships to more heavily embedded protostars requires the assumption that the dynamics of the underlying accretion process are similar across these evolutionary phases, and that the line luminosity can be measured accurately even when the protostar's photosphere is not detectable.  These assumptions are even more problematic for an extreme accretor such as PTF10nvg, but an accretion rate estimate via H~I line fluxes can still provide a useful lower limit on the object's accretion rate: changes in the accretion dynamics (i.e., boundary-layer accretion instead of magnetospheric accretion) or the presence of significant veiling flux will both likely result in an underestimate the object's full line flux, and/or an increase in the conversion factor between line luminosities and accretion rates.

We have generated crude mass accretion rate estimates for PTF10nvg from the strengths of its Br-$\gamma$ and Pa-$\beta$ emission lines.  We have dereddened the Pa-$\beta$ and Br-$\gamma$ line fluxes listed in Table \ref{tab:nir_lines} assuming an $A_V \approx 9$ mag (intermediate to the bounds calculated above), and converted those dereddened fluxes into line luminosities assuming a distance to PTF10nvg of $\sim$600 pc.  Using Equations 1 and 2 from \citet{Muzerolle1998} to estimate Log $L_{\rm acc}/{\rm L}_{\odot}$ from each line, we infer Log $L_{\rm acc}/{\rm L}_{\odot}$ = 0.62 and 0.335 from the Pa-$\beta$ and Br-$\gamma$ lines, respectively.  Assuming canonical Classical T Tauri Star parameters ($M = 0.5$~M$_{\odot}$ and $R = 3$~R$_{\odot}$), these accretion luminosities imply mass accretion rates of $\dot{M} \approx 2.5 \times 10^{-7}$ M$_{\odot}$ yr$^{-1}$. 

\subsubsection{Ca~II}\label{sec:caii}

The Ca~II triplet lines at 8498, 8542, and 8662~\AA\ are also generally associated with accretion activity in young stars.  PTF10nvg exhibits EWs for these lines of $-$15 to $-$25~\AA, varying among the observations, but in the realm of the rapidly accreting young Class II stars in Taurus and comparable to the emission levels measured by \citet{Connelley2010} in a large survey of Class I objects.  The line ratios seen from PTF10nvg varied, being somewhat typical of Class II sources ($F_{\rm Ca~II~8498}/F_{\rm Ca~II 8542} = 0.91$, $F_{\rm Ca~II~8662}/F_{\rm Ca~II~8542} = 0.81$) in the LRIS data taken on 2010 July 8, but roughly equal ($F_{\rm Ca~II~8498}/F_{\rm Ca~II 8542} = 1.08$, $F_{\rm Ca~II~8662}/F_{\rm Ca~II~8542} = 0.93$) in the SpeX data from July 14. The latter case is again similar to heavily accreting T Tauri stars at the extreme of the Class II range illustrated by \citet{Hamann1992} and close to the ratio predicted by their highly optically thick ($\tau=1000$) slab models.  It is also similar the outbursting behavior of V1647 Ori \citep{Walter2004}.  

Like the Paschen lines, the Ca~II triplet lines appear slightly asymmetric with respect to their line centers, having somewhat more flux on the blueshifted sides of the lines than the red.  

The Ca~II K line is also in emission, though the H line is affected by blueshifted Balmer line absorption.

\subsubsection{Other Atomic Lines}

The only absorption features we detect in these optical spectra are blueshifted lines of \iontoo{H}{i}, \iontoo{Na}{i} D, and \iontoo{K}{i}, along with the O~I triplet at 7773~\AA, which appears centered at the systemic velocity.   The IR spectra, similarly, have only a single blueshifted line in absorption (the He~I $\lambda$ 10830 feature seen in Fig. \ref{fig:PTF10nvg_IRSpec} and discussed in \S \ref{sub_sec:winds}), so no spectral type can be derived for the outbursting object.   Instead of a stellar photosphere, the spectrum is dominated by the circumstellar signatures of accretion and outflow processes. 

In addition to the H~I and Ca~II lines discussed in \S \ref{sec:hi} and \S \ref{sec:caii}, the optical and NIR spectra of PTF10nvg show many emission lines characteristic of young accreting stars.  The optical spectra show several permitted and forbidden species, including \iontoo{O}{i}, \iontoo{Fe}{i}, \iontoo{Fe}{ii}, \iontoo{Ca}{i}, \iontoo{Ca}{ii}, [Fe~II], [Ca~II], and [S~II].  The high-dispersion Keck NIRSPEC $Y$-band spectrum shows low-excitation ($<$1--2 eV) \iontoo{Ti}{i} lines as well as higher excitation (6--8 eV) \iontoo{Si}{i} and \iontoo{C}{i} lines in abundance, identified from the line list presented by \cite{Sharon2010}.  A Y-band survey including many Class II and some Class I stars in Taurus, as well as other young stars \citep{Edwards2006, Fischer2008}, suggests that some extreme emission line objects display the higher excitation lines.  The \iontoo{Ti}{i} is quite rare, however, and is found in our experience only in the spectra of V1331 Cyg, ZCMa, and SVS 13 -- all embedded sources driving strong outflows.  The lower dispersion SpeX spectrum shows emission lines from many neutral species including \iontoo{Al}{i}, \iontoo{Ca}{i}, \iontoo{Fe}{i}, \iontoo{K}{i}, \iontoo{Mg}{i}, \iontoo{Na}{i}, \iontoo{O}{i}, \iontoo{Si}{i}, and \iontoo{Ti}{i} over the 1--2.5 $\mu$m region; many of these features are also visible in the NIR outburst spectrum of V1647 (shown for comparison in Figure~\ref{fig:PTF10nvg_IRSpec}).

\begin{figure*}
\epsscale{0.8}
\plotone{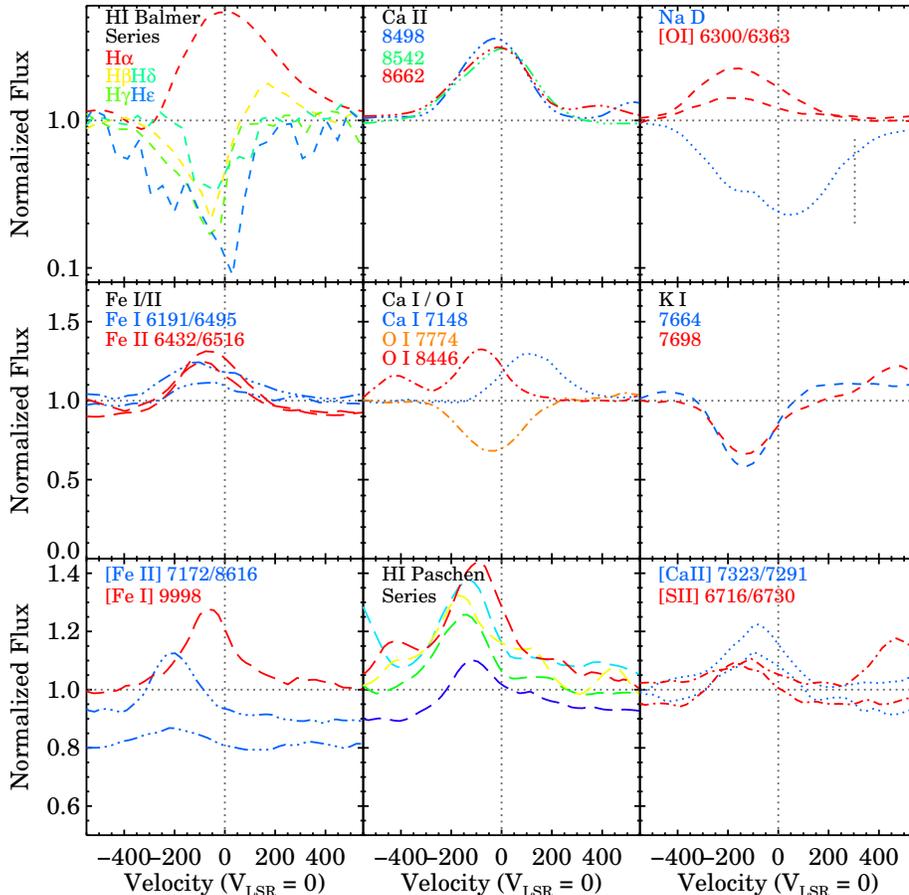}
\caption{ \normalsize{Selected spectral line profiles from the Keck LRIS spectrum, displayed with a velocity scale relative to the rest velocity of the Pelican Nebula ($V_{\rm LSR} \approx 0$ km s$^{-1}$).  }}\label{fig:PTF10nvg_OptSpec_kinematic}
\end{figure*}

\subsubsection{Wind Indicators: Blueshifted H~I, Na~D, K~I, \& He~I $\lambda$10830 Absorption and [O~I] Emission}\label{sub_sec:winds}

There is significant evidence for a substantial wind associated with the outburst of PTF10nvg.  Our optical spectra have limited resolution of $\sim$150--300 km s$^{-1}$, but nevertheless the H$\alpha$ and H$\beta$ lines display clear P-Cygni features with sub--continuum absorption out to $-400$ km s$^{-1}$ (see kinematic profiles shown in Figure \ref{fig:PTF10nvg_OptSpec_kinematic}).  The upper Balmer series lines lack the redshifted emission and display only a broad, blueshifted absorption trough.  In the NIR, where our NIRSPEC spectrum provides a higher resolution by a factor of 2--3, none of the Paschen or Brackett lines has P-Cygni structure.  The \iontoo{Na}{i} D lines, like the upper Balmer lines, are seen in blueshifted absorption out to about $-400$ km s$^{-1}$ but are blended at our resolution. 

The \iontoo{K}{i} doublet at 7664 and 7698~\AA\ is seen in absorption in PTF10nvg's spectrum, blueshifted by $\sim$175 km s$^{-1}$; this line is not typically seen in the spectra of Class I or Class II stars.  Strong and blueshifted \iontoo{K}{i} and \iontoo{Na}{i} D absorption, as well as P-Cygni Balmer profiles, are seen in the optical spectra of the FU Ori stars V1057 Cyg and V1515 Cyg, however, and in the outburst spectrum of V1647 Ori.  

\begin{figure*}
\epsscale{1.0}
\plotone{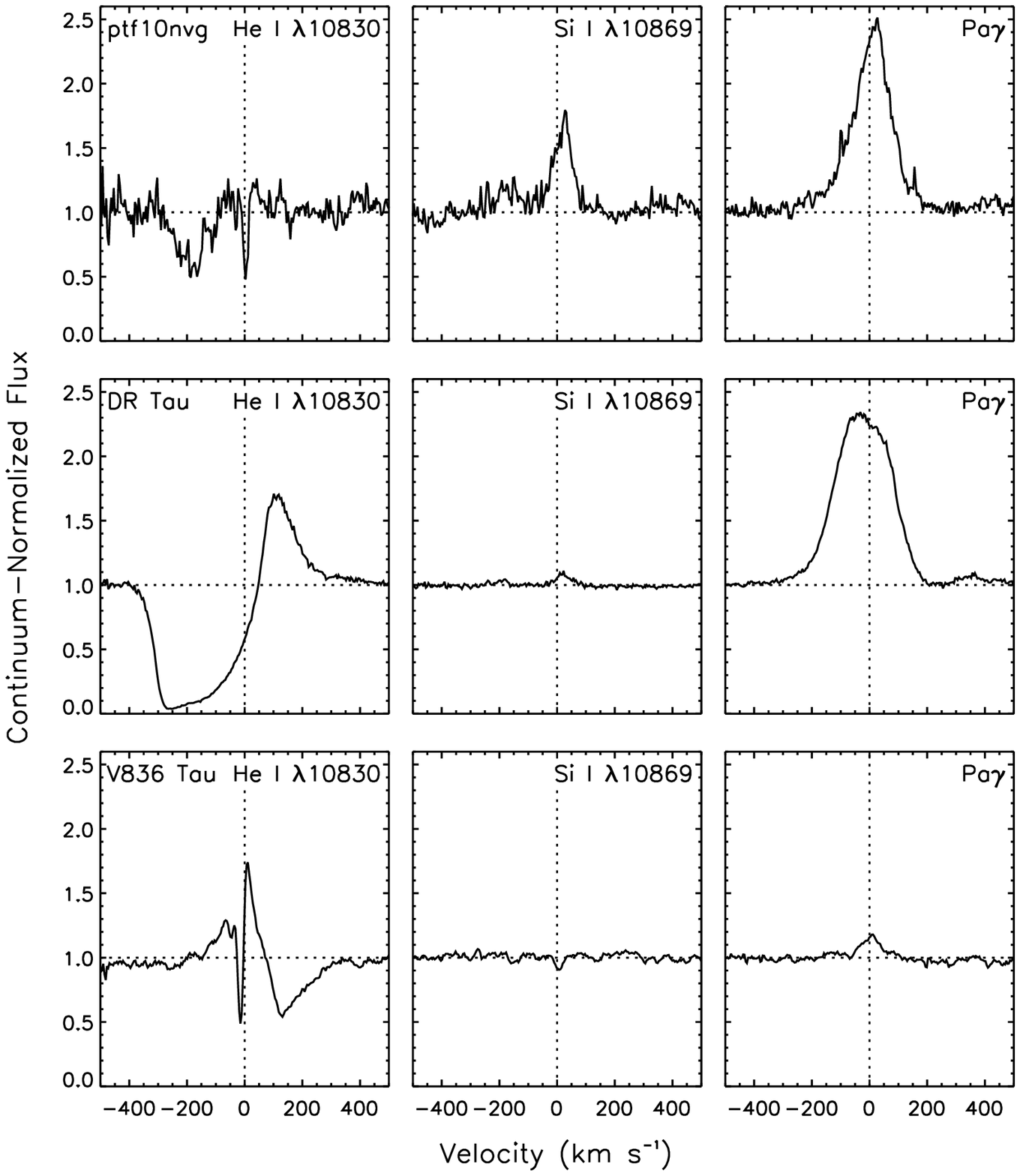}
\caption{ \normalsize{Keck NIRSPEC profiles of \iontoo{He}{i} $\lambda$10830, \iontoo{Si}{i} $\lambda$10869 in order to assess the possible
contamination of \iontoo{Si}{i} $\lambda$10830 to the He~I profile, and the H~I Pa-$\gamma$ line of PTF10nvg.  
DR Tau and V836 Tau are shown for comparison: the two Taurus stars are benchmarks for their comparable 
terminal velocity in the broad blueshifted absorption (DR Tau) and the roughly systemic narrow
absorption (V836 Tau) components to the \iontoo{He}{i} profile.
}}\label{fig:profilesY}
\end{figure*}

The high-dispersion NIRSPEC spectrum shows a somewhat complex He~I $\lambda$10830 profile (see Figure~\ref{fig:profilesY}).  Not unlike many heavily accreting T Tauri stars, such as DR Tau \citep{Edwards2003,Edwards2006}, there is strong blueshifted absorption out to about $-300$ km s$^{-1}$  which reaches 60\% of the continuum level. The absorption does not extend continuously inward to zero velocity, however: instead, the continuum level is reached by about $-100$ km s$^{-1}$ and there is a second narrow (FWHM $\approx 17$ km s$^{-1}$) absorption component centered at PTF10nvg's rest velocity.  Similar low-velocity and narrow absorption components are seen in the weakly or nonaccreting T Tauri stars V836 Tau and TWA 2.  When considering both the broad absorption terminal velocity and the narrow absorption depth, the \iontoo{He}{i} $\lambda$10830 profile of PTF10nvg is quite similar to that of V1057 Cyg (unpublished data).  The origin of a narrow absorption feature at roughly rest relative to the star is unknown; we note, however, the similar feature of V836 Tau was attributed to a disk wind by \cite{Kwan2007}.  Also of note is the possible presence of \iontoo{Si}{i} emission at 10830~\AA\ that could be filling in some of the blueshifted \iontoo{He}{i} absorption at intermediate velocities; this can be gauged by comparison to the \iontoo{Si}{i} $\lambda$10869 line also shown in Figure~\ref{fig:profilesY}.  

Other typical indicators of mass loss \citep{Hartigan1995} are the [O~I] $\lambda\lambda$6300, 6364 emission lines, which in PTF10nvg are symmetric and blueshifted by $\sim$175 km s$^{-1}$, and the [S~II] $\lambda\lambda$6717, 6731 lines, also blueshifted by $\sim$175 km s$^{-1}$.  The [N~II] $\lambda\lambda$6548, 6584 lines are relatively weak and [O~I] $\lambda$5577 is not apparent in our low spectral resolution data.  The forbidden-line EWs are comparable to those seen in the some of the strongest Class II stars in Taurus such as CW Tau, DD Tau, DG Tau, HN Tau, and UZ TauE  \citep{Hartigan1995}, but weaker than those typical of Class I and other stars in Taurus driving strong outflows \citep{White2004}.  

\citet{Hartigan1995} used detailed shock models to derive relations allowing a young star's mass loss rate to be estimated from the luminosities of its [O~I] 6300 and [S~II] 6731 forbidden lines.  These relations are based upon assumptions about the typical density, velocity, and abundance of the shock region, and also require that observed line fluxes be converted to line luminosities, with corrections to account for reddening.  Given the strong outflow signatures PTF10nvg possesses, it is possible, and perhaps likely, that the parameters commonly assumed for T Tauri star jets are not a good description of the physical properties of shocked regions in PTF10nvg's outflow.  Nonetheless, we can convert the observed [O~I] 6300 and [S~II] 6716 line fluxes to luminosities by adopting the same distance and extinction estimates used above to calculate HI line luminosities, and thus infer crude estimates of PTF10nvg's mass loss rate from Hartigan, Edwards \& Ghandour's (1995) equations A8 and A10.  These calculations imply estimated mass loss rates of 7 $\times$ 10$^{-7}$ M$_{\odot}$ yr$^{-1}$ (as inferred from [S~II] 6716) and 2 $\times$ 10$^{-6}$ M$_{\odot}$ yr$^{-1}$ (as inferred from [O~I] 6300).  Given the sizable uncertainties in the adopted extinction correction, and even the appropriateness of the physical properties assumed for the shock regions, these mass loss rates may easily be incorrect by more than an order of magnitude.  Nonetheless, they do emphasize the considerable mass loss associated with the PTF10nvg outburst; relative to the $\sim$10$^{-7}$ M$_{\odot}$ yr$^{-1}$ accretion rate inferred above from the HI emission lines, these estimates suggest an inflow/outflow ratio $\gtrsim$1, one to two orders of magnitude larger than the 0.01-0.1 ratios typically inferred for T Tauri stars.

NIR line ratios provide additional evidence that PTF10nvg is driving a strong outflow. \citet{Lorenzetti2009} and \citet{Antoniucci2008} have noted that EXor/jet driving sources typically demonstrate flux ratios between their Br-$\gamma$ and 2.206 $\mu$m \iontoo{Na}{i} emission lines of $\sim$ 2--5.  Class I sources that lack jet signatures, however, tend to have ratios that favor Br-$\gamma$ much more strongly \citep[e.g., 6--20; ][]{Nisini2005}.  The NIR spectra of PTF10nvg presented here possess a ratio of these lines of $\sim$2, placing PTF10nvg firmly in the company of the jet driving sources, as is consistent with the morphology of the wind-sensitive lines analyzed above.

\subsubsection{Outflow as Diagnosed via H$_2$ and [Fe~II] Emission}\label{sec:H2}

PTF10nvg shows clear emission in the 2.12 $\mu$m H$_2 (1-0)S(1)$ line, as well as a weaker feature which is likely the 2.24 $\mu$m $S(2-1)S(1)$ line.  Modeling has established that the ratio of these line strengths can diagnose between collisional and radiative excitation mechanisms: pioneering work by \citet{Gredel1995} demonstrated that UV and X-ray irradiation will produce distinctive H$_2 (1-0)S(1)$/$S(2-1)S(1)$ line ratios (0.54 and 0.06, respectively), and could also be distinguished from the ratio expected from a T=2000K shocked gas (0.13).  More recently, \citet{Nomura2007} calculated synthetic H$_2$ spectra as expected to emerge from protoplanetary disks with realistic gas-to-dust ratios and density/temperature profiles.  The \citet{Nomura2007} models predict that $H_2$ excitation by UV radiation is sensitive to the dust properties in the disk, producing H$_2 (1-0)S(1)$/$S(2-1)S(1)$ line ratios from 0.025--0.23 for power-law distributions of dust grains with maximum sizes of 10$\mu$m to 10 cm, respectively.  X-rays are much less sensitive to dust properties; models where X-rays dominate the excitation produce ratios of $\sim$0.07, irrespective of the dust properties.  

The H$_2 (1-0)S(1)$ and $S(2-1)S(1)$ line strengths recorded in Table \ref{tab:nir_lines} correspond to ratios of 0.23 and 0.3 for the SpeX and TSPEC observations; the less secure measurement of the (1-0)S(1) line suggest this ratio may best be treated as a tenuous detection which could be uncertain at the factor of two level.  Even accounting for this relatively large uncertainty, models attributing the excitation to X-rays alone are essentially ruled out; models incorporating UV or UV+X-ray heating and dust grains $\lesssim$1 mm would also be difficult to reconcile with this ratio.  The observed values are most consistent with, though slightly larger than, the \citet{Gredel1995} model of shock excitation or the \citet{Nomura2007} model of UV+X--ray irradiation of a disk with dust subject to coagulation and settling, which predict H$_2 (1-0)S(1)$ and $S(2-1)S(1)$ line ratios of 0.13 and $\sim$0.18, respectively.  Given the wealth of outflow signatures present in PTF10nvg's spectrum, we consider shocks to be the most likely source, but this conclusion is far from established.

[Fe~II] is another common outflow tracer for young stars: \citet{Connelley2010} demonstrate that the line strengths of H$_2$ and [Fe~II] are strongly correlated for a large sample of Class I protostars.  The TripleSpec observation of PTF10nvg shows clear [Fe~II] $\lambda$1.257 and $\lambda$1.644 $\mu$m emission; these lines may also be present in the SpeX observation, but the lower spectral resolution leads to a lower line-continuum contrast.  In comparison to the Class I stars in the \citet{Connelley2010} sample, PTF10nvg shows relatively weak H$_2$ and [Fe~II] emission.  \citet{Connelley2010} note, however, that the stars in their sample with weak H$_2$ also have low levels of continuum veiling, which is not true for PTF10nvg, where we cannot detect any photospheric absorption features.  

As both the 1.257 and 1.644 $\mu$m lines share the same upper level, \citet{Connelley2010} use their observed line fluxes to infer extinction estimates for each star in their sample, as with the H~I-based estimates presented in \S \ref{sec:hi}.  We find a line ratio in PTF10nvg of 1.48, essentially equal to that assumed by \citet{Connelley2010} for an unextinguished source, in strong disagreement with the extinction estimate of 6--12 which we infer from the H~I lines of PTF10nvg.  \citet{Connelley2010} find a similar discrepancy, however, in their analysis of the Class I source 06297+1021W, whose H~I lines imply $A_V \approx 18$ mag but whose [Fe~II] lines suggest $A_V \approx -1$ mag.  In many regards, 06297+1021W is quite similar to PTF10nvg: in addition to their similarly incongruous H~I and [Fe~II] extinction estimates, both show weak H$_2$ and [Fe~II] emission relative to other Class I sources \citep[][are unable to measure values for these features for 06297+1021W]{Connelley2010}, both objects appear as a $K$-band point source absent significant nebulosity, and both display a rich, morphologically similar NIR emission-line spectrum.  

\citet{Connelley2010} also explore the correlation between the strengths of the H$_2$, [Fe~II], and Br-$\gamma$ emission lines: relative to their sample of Class I objects, PTF10nvg lies on the weak side of these relations, though not significantly separated from the rest of the Class I population.

\subsubsection{Optical TiO/VO Emission}

In addition to the rich atomic emission-line spectrum, the most salient feature of the optical spectrum of PTF10nvg is the broad emission that we attribute to TiO and VO.  
The PTF10nvg spectrum (Figure ~\ref{fig:PTF10nvg_OptSpec})
looks much like an inverted mid-M spectral type star, suggesting that in addition to the hot and moderate density emission lines, there is a cooler \citep[$\sim$1500--4000 K][]{Lodders2002,Ferguson2005,Sharp2007} and high-density ($>10^{10}$ cm$^{-3}$) emitting component, perhaps the outer layers of a disk. 
Notably, the TiO emission is variable among our three spectra, being strongest just before the object brightened for the second time during the 2010 season.

To our knowledge, this is the first time such prominent hot molecular emission has been reported for a young star.   There is, however, literature discussion of emission in the optical TiO {\it bandheads}, though not the broader full bands as we observe in PTF10nvg.  TiO bandhead emission has been mentioned in the spectra of several luminous B[e] stars \citep{Zickgraf1989} and several red giant or supergiant dusty objects, notably VY CMa \citep{Phillips1987} and U Equ \citep{Barnbaum1996}, as well as in the red nova V4332 Sgr \citep{Goranskii2007}.   Most relevant to the present case is the report by \citet{Herbig2009} that V1057 Cyg exhibited emission in 2004 in ``the heads of the (0-0) bands of the $\gamma$ system of TiO at 7054, 7087, and 7125 \AA."  We have confirmed this narrow emission in our own data on V1057 Cyg but emphasize that the PTF10nvg TiO emission is through the full band region, not just in the bandheads.  

\subsubsection{Infrared CO, VO, and H$_2$O Emission Lines}

In addition to the optical TiO and VO emission, there is also molecular emission throughout the 1--5 $\mu$m spectra of PTF10nvg. Readily apparent broad features above the likely dust-dominated red continuum are VO at 1.05 $\mu$m, several H$_2$0 bands around 1.4, 1.8, 2.0, and 2.45 $\mu$m as well as the Q and R branches at 3.0 and 3.3 $\mu$m, and prominent CO first overtone bandheads beginning at 2.3 $\mu$m.  While the VO and H$_2$O suggest emitting region temperatures of $< 3500$~K, the CO emission could come from warmer (2500--5000~K) dense ($>10^{10}$ cm$^{-3}$) gas.

\citet{Connelley2010} display correlations between the morphology and strengths of the CO, Na, Ca, and Br-$\gamma$ features for Class I sources, finding that all sources with CO in emission also display Br-$\gamma$ and Na in emission.  PTF10nvg follows this relation, and displays CO, Na, and Ca emission-line strengths similar to those of other Class I sources with each set of features in emission.  PTF10nvg does show somewhat weaker Br-$\gamma$ relative to its CO emission strength than most Class I sources, but is not an extreme outlier to that relation.  

In contrast to the CO bandhead emission at 2.3 $\mu$m, the 4.7 $\mu$m CO fundamental band is in {\it absorption} with both the P and R branches evident.

\section{Discussion}

The spectra obtained to date of PTF10nvg lack the strong FG-supergiant optical and M-supergiant NIR absorption features typical of FU Ori objects, which are interpreted as disk-atmosphere dominated systems.    Instead, the PTF10nvg spectra are characterized mainly by emission lines, some with kinematic signatures of formation in outflowing material. The only similarities to the FU Ori class are in the \iontoo{Na}{i} D, \iontoo{K}{i}, and \iontoo{He}{i} blueshifted absorptions which are kinematically associated with strong outflows, as well as the \iontoo{O}{i} 7773 \AA\ neutral velocity absorption which also may be related to winds. Despite the optical rise by over 6 mag within one year, PTF10nvg does not appear to be in the FU Ori class.  This statement is based on the first year of data on this object and does not preclude that it could settle down in terms of its variability and its outflow-dominated optical and NIR spectrum to reveal FU Ori-like absorption features in the future.  There is, however, a notable parallel to the FU Ori object V1057 Cyg in the He~I $\lambda$10830 profile and a similarity in the optical TiO emission.  

The similarities of PTF10nvg to EX Lup outbursts are also slim. Neither the light curve nor the spectra are similar to this class of objects, which are interpreted as episodes of enhanced accretion but less extreme than those nominally driven by disk instability in FU Ori objects.  

Instead, the PTF10nvg data bear qualitative and quantitative similarities to the 2004 outburst of V1647 Ori (also known as McNeil's Nebula), albeit with a significantly weaker wind component.  Specifically, the amplitude of the optical outburst is similar, although PTF10nvg is much more unstable and photometrically variable in its outburst phase (see the $I$-band light curve given by \citet{Acosta-Pulido2007}.

The PTF10nvg SpeX spectrum is a reasonable match to the 1 year post-outburst spectrum of V1647 Ori in terms of the emission-line features.  The most notable disagreement is the absence of broad 3.3 and 4.7 $\mu$m ice features in the spectrum of PTF10nvg: those features are present at all epochs for V1647 Ori and thought to be foreground molecular cloud material not related to the outburst.  The early V1647 Ori spectrum (March 2004) also showed P-Cygni profiles in Paschen lines, not seen in PTF10nvg, but those profiles disappeared by March 2005 \citep[e.g.,][]{Gibb2006}.  Along the same lines, compared to only 1.083 $\mu$m He I visible in PTF10nvg, V1647 Ori showed strong He I absorption at both 1.083 and 2.05 $\mu$m during its recent outburst, with slight red-side emission in both lines \citep[Fig. 1, ][]{Vacca2004}, but by Nov. 2004 the 1.083 $\mu$m line was much weaker, without red-side emission \citep[Fig. 1, ][]{Gibb2006}.  Notably, despite the strong signatures of outflow there is no evidence of shocked H$_2$ emission in V1647 Ori.  PTF10nvg, in contrast, has a relatively slow wind but does possess H$_2$ in emission.  This could be a line-of-sight effect but may be an intrinsic astrophysical difference.  There are several objects in the \citet{Connelley2010} catalog of Class I sources which show somewhat similar NIR spectra to PTF10nvg and also display H$_2$ in emission: 16289-4449 (another eruptive variable with CO in emission), and 20453+6746, whose spectrum is somewhat more in line with a typical Class I source but which does display some of the same emission lines in the $J$ band.  06297+1021W, the Class I source in the \citet{Connelley2010} catalog whose similarities to PTF10nvg were outlined in \S \ref{sec:H2}, does not, however, appear to show strong H$_2$ emission.  

The PTF10nvg optical spectra also appear similar to those of V1647 Ori in terms of line EWs and flux ratios, though they indicate weaker P-Cygni structure in the Balmer and Ca~II lines \citep[e.g.,][]{Reipurth2004,Briceno2004}, which themselves disappeared in V1647 Ori roughly 1 year post-outburst \citep{Ojha2006}.

A unique feature, however, of PTF10nvg relative to any known young stellar object is the molecular TiO and VO seen in emission in the optical spectra.  This emission has varied in strength relative to the varying continuum over a 1 month time scale.  It indicates the presense of dense warm gas, perhaps the upper levels of a disk atmosphere heated by irradiation from enhanced accretion luminosity.

\section{Summary}

\begin{enumerate}
\item{We have identified PTF10nvg, a Class I protostar undergoing a strong outburst.  A comparison of the source's pre- and post-outburst SED suggests that the bolometric luminosity has increased by a factor of more than 100.}
\item{This outburst is presumably driven by a significant increase in PTF10nvg's mass accretion rate. We analyzed the strength of the NIR H~I lines to derive an extinction estimate of $A_V \sim$ 6--12 mag, and an accretion rate of $\dot{M} \approx$ 2.5 $\times 10^{-7}$ M$_{\odot}$ yr$^{-1}$.}  
\item{Several wind-sensitive optical and NIR absorption lines exhibit blueshifts of several hundred km s$^{-1}$, indicating that PTF10nvg is driving a substantial outflow.  This interpretation is supported by the presence of several spectroscopic lines thought to trace shocked gas (H$_{2}$, [S~II], [Fe~II]). }
\item{Several optical TiO bands are seen fully in emission, a characteristic that is, to the best of our knowledge, unique to this astronomical object. This emission reveals the presence of a significant amount of dense ($n > 10^{10}$ cm$^{-3}$), warm (1500--4000 K), presumably circumstellar gas.}
\item{During its present outburst, PTF10nvg's photometric and spectroscopic properties differ significantly from those commonly associated with EXor or FU Ori outbursts.  We do identify several objects whose NIR spectra show morphological similarities to that of PTF10nvg: 06297+1021W, 16289-4449, and 20453+6746, as well as V1647 Ori, at least during its 2004--2006 outburst.  Detailed monitoring of PTF10nvg, and each of these other variables, is required to ascertain the exact nature of these enigmatic sources.}
\end{enumerate}

\acknowledgements The authors are grateful to Michael T. Kandrashoff, Christopher V. Griffith, A. Cucchiara, Daniel A. Perley, and A. N. Morgan for their assistance in obtaining the LRIS and Kast spectra presented here.  We thank the Directors of the NASA Infrared Telescope Facility (IRTF) and Apache Point Observatory (APO), whose allocation of Director's Discretionary Time enabled the acquisition of the SpeX and TripleSpec spectra presented here, and an anonymous referee, whose prompt and helpful review improved the content and presentation of this work.  We also thank Erika Gibb for making available archival SpeX observations of V1647 Ori, Tom Greene for a helpful discussion of NIR shock tracers, and Michael Eracleous for a useful discussion of symbiotic binaries and other accreting systems.  

K.R.C. acknowledges support for this work from the Hubble Fellowship Program, provided by NASA through Hubble Fellowship grant HST-HF-51253.01-A awarded by the STScI, which is operated by the AURA, Inc., for NASA, under contract NAS 5-26555. J.S.B., D.A.P., C.K., A.A.M., and D.A.S. acknowledge support of an NSF-CDI grant, ``Real-Time Classification of Massive Time-Series Data Streams" (Award \#0941742).  A.V.F.'s group is grateful for the support of NSF grant AST-0908886, the TABASGO Foundation, Gary and Cynthia Bengier, and the Richard and Rhoda Goldman Fund.  

Some of the data presented herein were obtained at the W. M. Keck Observatory, which is operated as a scientific partnership among the California Institute of Technology, the University of California, and the National Aeronautics and Space Administration. The Observatory was made possible by the generous financial support of the W. M. Keck Foundation. The authors wish to recognize and acknowledge the very significant cultural role and reverence that the summit of Mauna Kea has always had within the indigenous Hawaiian community.  We are most fortunate to have the opportunity to conduct observations from this mountain.

PAIRITEL is operated by the Smithsonian Astrophysical Observatory (SAO) and was made possible by a grant from the Harvard University Milton Fund, a camera loan from the University of Virginia, and continued support of the SAO and UC Berkeley.  The PAIRITEL project and those working on PAIRITEL data are further supported by NASA/Swift Guest Investigator Programs NNX09AQ66Q and NNX10A128G.
This work was also based in part on observations obtained with the Apache Point Observatory 3.5~m telescope, which is owned and operated by the Astrophysical Research Consortium. We are grateful for the assistance of the staffs at all of the observatories used to obtain the data.

This research has made use of NASA's Astrophysics Data System
Bibliographic Services, the SIMBAD database, operated at CDS,
Strasbourg, France, the NASA/IPAC Extragalactic Database, operated by
the Jet Propulsion Laboratory, California Institute of Technology,
under contract with the National Aeronautics and Space Administration,
and the VizieR database of astronomical catalogs
\citep{Ochsenbein2000}. 

This publication makes use of data products from the Two Micron All Sky Survey, which is a joint project of the University of Massachusetts and the Infrared Processing and Analysis Center/California Institute of Technology, funded by the National Aeronautics and Space Administration and the National Science Foundation. 

The National Energy Research Scientific Computing Center, which is supported by the Office of Science of the U.S. Department of Energy under Contract No. DE-AC02-05CH11231, provided staff, computational resources, and data storage for this project. 

\facility{Palomar: 48 inch (PTF)}
\facility{Palomar: 200 inch (DBSP)}
\facility{Keck (LRIS, NIRSPEC)}
\facility{Lick:3 m (Kast)}
\facility{IRTF (SpeX)}
\facility{Apache Point Observatory (TripleSpec)}
\facility{FLWO: 1.3 m (PAIRITEL)}

\setlength{\baselineskip}{0.6\baselineskip}

\begin{thebibliography}{86}
\expandafter\ifx\csname natexlab\endcsname\relax\def\natexlab#1{#1}\fi

\bibitem[{{Acosta-Pulido} {et~al.}(2007){Acosta-Pulido}, {Kun},
  {{\'A}brah{\'a}m}, {K{\'o}sp{\'a}l}, {Csizmadia}, {Kiss}, {Mo{\'o}r},
  {Szabados}, {Benk{\H o}}, {Barrena Delgado}, {Charcos-Llorens}, {Eredics},
  {Kiss}, {Manchado}, {R{\'a}cz}, {Ramos Almeida}, {Sz{\'e}kely}, \&
  {Vidal-N{\'u}{\~n}ez}}]{Acosta-Pulido2007}
{Acosta-Pulido}, J.~A. {et~al.} 2007, \aj, 133, 2020

\bibitem[{{Antoniucci} {et~al.}(2008){Antoniucci}, {Nisini}, {Giannini}, \&
  {Lorenzetti}}]{Antoniucci2008}
{Antoniucci}, S., {Nisini}, B., {Giannini}, T., \& {Lorenzetti}, D. 2008, \aap,
  479, 503

\bibitem[{{Armitage} {et~al.}(2001){Armitage}, {Livio}, \&
  {Pringle}}]{Armitage2001}
{Armitage}, P.~J., {Livio}, M., \& {Pringle}, J.~E. 2001, \mnras, 324, 705

\bibitem[{{Aspin} {et~al.}(2009){Aspin}, {Reipurth}, {Beck}, {Aldering},
  {Doering}, {Hammel}, {Lynch}, {Meixner}, {Pecontal}, {Russell}, {Sitko},
  {Thomas}, \& {U}}]{Aspin2009}
{Aspin}, C. {et~al.} 2009, \apjl, 692, L67

\bibitem[{{Astier} {et~al.}(2006){Astier}, {Guy}, {Regnault}, {Pain},
  {Aubourg}, {Balam}, {Basa}, {Carlberg}, {Fabbro}, {Fouchez}, {Hook},
  {Howell}, {Lafoux}, {Neill}, {Palanque-Delabrouille}, {Perrett}, {Pritchet},
  {Rich}, {Sullivan}, {Taillet}, {Aldering}, {Antilogus}, {Arsenijevic},
  {Balland}, {Baumont}, {Bronder}, {Courtois}, {Ellis}, {Filiol}, {Gon{\c
  c}alves}, {Goobar}, {Guide}, {Hardin}, {Lusset}, {Lidman}, {McMahon},
  {Mouchet}, {Mourao}, {Perlmutter}, {Ripoche}, {Tao}, \&
  {Walton}}]{Astier2006}
{Astier}, P. {et~al.} 2006, \aap, 447, 31

\bibitem[{{Audard} {et~al.}(2010){Audard}, {Stringfellow}, {G{\"u}del},
  {Skinner}, {Walter}, {Guinan}, {Hamilton}, {Briggs}, \&
  {Baldovin-Saavedra}}]{Audard2010}
{Audard}, M. {et~al.} 2010, \aap, 511, A63

\bibitem[{{Bally} \& {Reipurth}(2003)}]{Bally2003}
{Bally}, J., \& {Reipurth}, B. 2003, \aj, 126, 893

\bibitem[Barnbaum et 
al.(1996)]{Barnbaum1996} Barnbaum, C., Omont, A., \& Morris, M.\ 1996, \aap, 310, 259 

\bibitem[{{Bell} \& {Lin}(1994)}]{Bell1994}
{Bell}, K.~R., \& {Lin}, D.~N.~C. 1994, \apj, 427, 987

\bibitem[Bertin \& Arnouts(1996)]{Bertin1996} Bertin, E., \& Arnouts, S.\ 1996, \aaps, 117, 393 

\bibitem[{{Bessell}(1999)}]{b99}
{Bessell}, M.~S. 1999, \pasp, 111, 1426

\bibitem[Blake et al.(2008)]{Blake2008} Blake, C.~H., Bloom, 
J.~S., Latham, D.~W., Szentgyorgyi, A.~H., Skrutskie, M.~F., Falco, E.~E., 
\& Starr, D.~S.\ 2008, \pasp, 120, 860 

\bibitem[{{Bloom} {et~al.}(2006){Bloom}, {Starr}, {Blake}, {Skrutskie}, \&
  {Falco}}]{Bloom2006}
{Bloom}, J.~S., {Starr}, D.~L., {Blake}, C.~H., {Skrutskie}, M.~F., \& {Falco},
  E.~E. 2006, in Astronomical Society of the Pacific Conference Series, Vol.
  351, Astronomical Data Analysis Software and Systems XV, ed. {C.~Gabriel,
  C.~Arviset, D.~Ponz, \& S.~Enrique}, 751

\bibitem[Bloom et al.(2009)]{Bloom2009} Bloom, J.~S., et al.\ 
2009, \apj, 691, 723 

\bibitem[{{Bochanski} {et~al.}(2007){Bochanski}, {West}, {Hawley}, \&
  {Covey}}]{Bochanski2007}
{Bochanski}, J.~J., {West}, A.~A., {Hawley}, S.~L., \& {Covey}, K.~R. 2007,
  \aj, 133, 531

\bibitem[{{Boley} {et~al.}(2006){Boley}, {Mej{\'{\i}}a}, {Durisen}, {Cai},
  {Pickett}, \& {D'Alessio}}]{Boley2006}
{Boley}, A.~C., {Mej{\'{\i}}a}, A.~C., {Durisen}, R.~H., {Cai}, K., {Pickett},
  M.~K., \& {D'Alessio}, P. 2006, \apj, 651, 517

\bibitem[{{Botticella} {et~al.}(2010){Botticella}, {Trundle}, {Pastorello},
  {Rodney}, {Rest}, {Gezari}, {Smartt}, {Narayan}, {Huber}, {Tonry}, {Young},
  {Smith}, {Bresolin}, {Valenti}, {Kotak}, {Mattila}, {Kankare}, {Wood-Vasey},
  {Riess}, {Neill}, {Forster}, {Martin}, {Stubbs}, {Burgett}, {Chambers},
  {Dombeck}, {Flewelling}, {Grav}, {Heasley}, {Hodapp}, {Kaiser}, {Kudritzki},
  {Luppino}, {Lupton}, {Magnier}, {Monet}, {Morgan}, {Onaka}, {Price},
  {Rhoads}, {Siegmund}, {Sweeney}, {Wainscoat}, {Waters}, {Waterson}, \&
  {Wynn-Williams}}]{Botticella2010}
{Botticella}, M.~T. {et~al.} 2010, \apjl, 717, L52

\bibitem[{{Brice{\~n}o} {et~al.}(2004){Brice{\~n}o}, {Vivas}, {Hern{\'a}ndez},
  {Calvet}, {Hartmann}, {Megeath}, {Berlind}, {Calkins}, \&
  {Hoyer}}]{Briceno2004}
{Brice{\~n}o}, C. {et~al.} 2004, \apjl, 606, L123

\bibitem[Burrows et al.(2005)]{Burrows2005} Burrows, D.~N., et al.\ 
2005, \ssr, 120, 165 

\bibitem[Butler 
\& Kocevski(2007)]{Butler2007} Butler, N.~R., \& Kocevski, D.\ 2007, \apj, 668, 400 

\bibitem[{{Carpenter}(2001)}]{Carpenter2001}
{Carpenter}, J.~M. 2001, \aj, 121, 2851

\bibitem[{{Cenko} {et~al.}(2006){Cenko}, {Fox}, {Moon}, {Harrison}, {Kulkarni},
  {Henning}, {Guzman}, {Bonati}, {Smith}, {Thicksten}, {Doyle}, {Petrie},
  {Gal-Yam}, {Soderberg}, {Anagnostou}, \& {Laity}}]{Cenko2006}
{Cenko}, S.~B. {et~al.} 2006, \pasp, 118, 1396

\bibitem[{{Connelley} \& {Greene}(2010)}]{Connelley2010}
{Connelley}, M.~S., \& {Greene}, T.~P. 2010, ArXiv e-prints

\bibitem[{{Covey} {et~al.}(2007){Covey}, {Ivezi{\'c}}, {Schlegel},
  {Finkbeiner}, {Padmanabhan}, {Lupton}, {Ag{\"u}eros}, {Bochanski}, {Hawley},
  {West}, {Seth}, {Kimball}, {Gogarten}, {Claire}, {Haggard}, {Kaib},
  {Schneider}, \& {Sesar}}]{Covey2007}
Covey, K. et~al. 2007, \aj, 134, 2398

\bibitem[{{Cushing} {et~al.}(2004){Cushing}, {Vacca}, \&
  {Rayner}}]{Cushing2004}
{Cushing}, M.~C., {Vacca}, W.~D., \& {Rayner}, J.~T. 2004, \pasp, 116, 362

\bibitem[Dobashi et al.(1994)]{Dobashi1994} Dobashi, K., Bernard, 
J.-P., Yonekura, Y., \& Fukui, Y.\ 1994, \apjs, 95, 419 

\bibitem[{{Edwards} {et~al.}(2006){Edwards}, {Fischer}, {Hillenbrand}, \&
  {Kwan}}]{Edwards2006}
{Edwards}, S., {Fischer}, W., {Hillenbrand}, L., \& {Kwan}, J. 2006, \apj, 646,
  319

\bibitem[{{Edwards} {et~al.}(2003){Edwards}, {Fischer}, {Kwan}, {Hillenbrand},
  \& {Dupree}}]{Edwards2003}
{Edwards}, S., {Fischer}, W., {Kwan}, J., {Hillenbrand}, L., \& {Dupree}, A.~K.
  2003, \apjl, 599, L41

\bibitem[Egan et al.(1999)]{Egan1999} Egan, M.~P., Price, S.~D., 
Shipman, R.~F., Gugliotti, G.~M., Tedesco, E.~F., Moshir, M., 
\& Cohen, M.\ 1999, Astrophysics with Infrared Surveys: A Prelude to SIRTF, 177, 404 

\bibitem[Enoch et al.(2009)]{Enoch2009} Enoch, M.~L., Evans, 
N.~J., Sargent, A.~I., \& Glenn, J.\ 2009, \apj, 692, 973 

\bibitem[{{Fedele} {et~al.}(2007){Fedele}, {van den Ancker}, {Petr-Gotzens}, \&
  {Rafanelli}}]{Fedele2007}
{Fedele}, D., {van den Ancker}, M.~E., {Petr-Gotzens}, M.~G., \& {Rafanelli},
  P. 2007, \aap, 472, 207

\bibitem[Ferguson et al.(2005)]{Ferguson2005} Ferguson, J.~W., 
Alexander, D.~R., Allard, F., Barman, T., Bodnarik, J.~G., Hauschildt, 
P.~H., Heffner-Wong, A., \& Tamanai, A.\ 2005, \apj, 623, 585 

\bibitem[{{Filippenko}(1982)}]{Filippenko1982}
{Filippenko}, A.~V. 1982, \pasp, 94, 715

\bibitem[Finkbeiner et al.(2004)]{Finkbeiner2004} Finkbeiner, D.~P., 
et al.\ 2004, \aj, 128, 2577 

\bibitem[Fischer et al.(2008)]{Fischer2008} Fischer, W., Kwan, J., 
Edwards, S., \& Hillenbrand, L.\ 2008, \apj, 687, 1117 

\bibitem[{{Gibb} {et~al.}(2006){Gibb}, {Rettig}, {Brittain}, {Wasikowski},
  {Simon}, {Vacca}, {Cushing}, \& {Kulesa}}]{Gibb2006}
{Gibb}, E.~L., {Rettig}, T.~W., {Brittain}, S.~D., {Wasikowski}, D., {Simon},
  T., {Vacca}, W.~D., {Cushing}, M.~C., \& {Kulesa}, C. 2006, \apj, 641, 383

\bibitem[Goranskii 
\& Barsukova(2007)]{Goranskii2007} Goranskii, V.~P., \& Barsukova, E.~A.\ 2007, Astronomy Reports, 51, 126 

\bibitem[{{Gredel} \& {Dalgarno}(1995)}]{Gredel1995}
{Gredel}, R., \& {Dalgarno}, A. 1995, \apj, 446, 852

\bibitem[Greene et al.(2008)]{Greene2008} Greene, T.~P., Aspin, 
C., \& Reipurth, B.\ 2008, \aj, 135, 1421 

\bibitem[{{Guieu} {et~al.}(2009){Guieu}, {Rebull}, {Stauffer}, {Hillenbrand},
  {Carpenter}, {Noriega-Crespo}, {Padgett}, {Cole}, {Carey}, {Stapelfeldt}, \&
  {Strom}}]{Guieu2009}
{Guieu}, S. {et~al.} 2009, \apj, 697, 787

\bibitem[{{Hamann} \& {Persson}(1992)}]{Hamann1992}
{Hamann}, F., \& {Persson}, S.~E. 1992, \apjs, 82, 247

\bibitem[{{Hartigan} {et~al.}(1995){Hartigan}, {Edwards}, \&
  {Ghandour}}]{Hartigan1995}
{Hartigan}, P., {Edwards}, S., \& {Ghandour}, L. 1995, \apj, 452, 736

\bibitem[{{Hartmann} \& {Kenyon}(1996)}]{Hartmann1996}
{Hartmann}, L., \& {Kenyon}, S.~J. 1996, \araa, 34, 207

\bibitem[{{Herbig}(1958)}]{Herbig1958}
{Herbig}, G.~H. 1958, \apj, 128, 259

\bibitem[{{Herbig}(1989)}]{Herbig1989}
{Herbig}, G.~H. 1989, in European Southern Observatory Conference and Workshop
  Proceedings, Vol.~33, European Southern Observatory Conference and Workshop
  Proceedings, ed. {B.~Reipurth}, 233

\bibitem[{{Herbig}(2008)}]{Herbig2008}
---. 2008, \aj, 135, 637

\bibitem[{{Herbig}(2009)}]{Herbig2009}
---. 2009, \aj, 138, 448

\bibitem[{{Herbig} {et~al.}(2001){Herbig}, {Aspin}, {Gilmore}, {Imhoff}, \&
  {Jones}}]{Herbig2001}
{Herbig}, G.~H., {Aspin}, C., {Gilmore}, A.~C., {Imhoff}, C.~L., \& {Jones},
  A.~F. 2001, \pasp, 113, 1547

\bibitem[{{Horne}(1986)}]{h86}
{Horne}, K. 1986, \pasp, 98, 609

\bibitem[Ishihara et al.(2010)]{Ishihara2010} Ishihara, D., et al.\ 2010, \aap, 514, A1 

\bibitem[{{Joy}(1945)}]{Joy1945}
{Joy}, A.~H. 1945, \apj, 102, 168

\bibitem[Kastner et al.(2006)]{Kastner2006} Kastner, J.~H., et al.\ 
2006, \apjl, 648, L43 

\bibitem[{{Kley} \& {Lin}(1999)}]{Kley1999}
{Kley}, W., \& {Lin}, D.~N.~C. 1999, \apj, 518, 833

\bibitem[{{Kwan} {et~al.}(2007){Kwan}, {Edwards}, \& {Fischer}}]{Kwan2007}
{Kwan}, J., {Edwards}, S., \& {Fischer}, W. 2007, \apj, 657, 897

\bibitem[Lada(1987)]{Lada1987} Lada, C.~J.\ 1987, Star Forming 
Regions, 115, 1 

\bibitem[{{Law} {et~al.}(2009){Law}, {Kulkarni}, {Dekany}, {Ofek}, {Quimby},
  {Nugent}, {Surace}, {Grillmair}, {Bloom}, {Kasliwal}, {Bildsten}, {Brown},
  {Cenko}, {Ciardi}, {Croner}, {Djorgovski}, {van Eyken}, {Filippenko}, {Fox},
  {Gal-Yam}, {Hale}, {Hamam}, {Helou}, {Henning}, {Howell}, {Jacobsen},
  {Laher}, {Mattingly}, {McKenna}, {Pickles}, {Poznanski}, {Rahmer}, {Rau},
  {Rosing}, {Shara}, {Smith}, {Starr}, {Sullivan}, {Velur}, {Walters}, \&
  {Zolkower}}]{Law2009}
{Law}, N.~M. {et~al.} 2009, \pasp, 121, 1395

\bibitem[{{Lodato} \& {Clarke}(2004)}]{Lodato2004}
{Lodato}, G., \& {Clarke}, C.~J. 2004, \mnras, 353, 841

\bibitem[Lodders(2002)]{Lodders2002} Lodders, K.\ 2002, \apj, 577, 
974 

\bibitem[{{Lorenzetti} {et~al.}(2007){Lorenzetti}, {Giannini}, {Larionov},
  {Kopatskaya}, {Arkharov}, {De Luca}, \& {Di Paola}}]{Lorenzetti2007}
{Lorenzetti}, D., {Giannini}, T., {Larionov}, V.~M., {Kopatskaya}, E.,
  {Arkharov}, A.~A., {De Luca}, M., \& {Di Paola}, A. 2007, \apj, 665, 1182

\bibitem[{{Lorenzetti} {et~al.}(2009){Lorenzetti}, {Larionov}, {Giannini},
  {Arkharov}, {Antoniucci}, {Nisini}, \& {Di Paola}}]{Lorenzetti2009}
{Lorenzetti}, D., {Larionov}, V.~M., {Giannini}, T., {Arkharov}, A.~A.,
  {Antoniucci}, S., {Nisini}, B., \& {Di Paola}, A. 2009, \apj, 693, 1056

\bibitem[{{Magnier} {et~al.}(1999){Magnier}, {Volp}, {Laan}, {van den Ancker},
  \& {Waters}}]{Magnier1999}
{Magnier}, E.~A., {Volp}, A.~W., {Laan}, K., {van den Ancker}, M.~E., \&
  {Waters}, L.~B.~F.~M. 1999, \aap, 352, 228

\bibitem[{{Matheson} {et~al.}(2000){Matheson}, {Filippenko}, {Ho}, {Barth}, \&
  {Leonard}}]{mfh+00}
{Matheson}, T., {Filippenko}, A.~V., {Ho}, L.~C., {Barth}, A.~J., \& {Leonard},
  D.~C. 2000, \aj, 120, 1499

\bibitem[{{McCarthy} {et~al.}(1998){McCarthy}, {Cohen}, {Butcher}, {Cromer},
  {Croner}, {Douglas}, {Goeden}, {Grewal}, {Lu}, {Petrie}, {Weng}, {Weber},
  {Koch}, \& {Rodgers}}]{McCarthy1998}
{McCarthy}, J.~K. {et~al.} 1998, in Society of Photo-Optical Instrumentation
  Engineers (SPIE) Conference Series, Vol. 3355, Society of Photo-Optical
  Instrumentation Engineers (SPIE) Conference Series, ed. {S.~D'Odorico},
  81

\bibitem[{{McNeil} {et~al.}(2004){McNeil}, {Reipurth}, \& {Meech}}]{McNeil2004}
{McNeil}, J.~W., {Reipurth}, B., \& {Meech}, K. 2004, \iaucirc, 8284, 1

\bibitem[{{Meyer} {et~al.}(1997){Meyer}, {Calvet}, \&
  {Hillenbrand}}]{Meyer1997}
{Meyer}, M.~R., {Calvet}, N., \& {Hillenbrand}, L.~A. 1997, \aj, 114, 288

\bibitem[{{Miller} \& {Stone}(1993)}]{Miller1993}
{Miller}, J.~S., \& {Stone}, R.~P.~S. 1993, {Lick Obs. Tech. Rep. 66} (Santa
  Cruz: Lick Obs.)

\bibitem[{{Monet} {et~al.}(2003){Monet}, {Levine}, {Canzian}, {Ables}, {Bird},
  {Dahn}, {Guetter}, {Harris}, {Henden}, {Leggett}, {Levison}, {Luginbuhl},
  {Martini}, {Monet}, {Munn}, {Pier}, {Rhodes}, {Riepe}, {Sell}, {Stone},
  {Vrba}, {Walker}, {Westerhout}, {Brucato}, {Reid}, {Schoening}, {Hartley},
  {Read}, \& {Tritton}}]{Monet2003}
{Monet}, D.~G. {et~al.} 2003, \aj, 125, 984

\bibitem[{{Morales-Calder{\'o}n} {et~al.}(2009){Morales-Calder{\'o}n},
  {Stauffer}, {Rebull}, {Whitney}, {Barrado y Navascu{\'e}s}, {Ardila}, {Song},
  {Brooke}, {Hartmann}, \& {Calvet}}]{Morales-Calderon2009}
{Morales-Calder{\'o}n}, M. {et~al.} 2009, \apj, 702, 1507

\bibitem[{{Muzerolle} {et~al.}(1998){Muzerolle}, {Hartmann}, \&
  {Calvet}}]{Muzerolle1998}
{Muzerolle}, J., {Hartmann}, L., \& {Calvet}, N. 1998, \aj, 116, 2965

\bibitem[{{Nisini} {et~al.}(2005){Nisini}, {Antoniucci}, {Giannini}, \&
  {Lorenzetti}}]{Nisini2005}
{Nisini}, B., {Antoniucci}, S., {Giannini}, T., \& {Lorenzetti}, D. 2005, \aap,
  429, 543

\bibitem[Nomura et al.(2007)]{Nomura2007} Nomura, H., Aikawa, Y., 
Tsujimoto, M., Nakagawa, Y., \& Millar, T.~J.\ 2007, \apj, 661, 334 

\bibitem[{{Ochsenbein} {et~al.}(2000){Ochsenbein}, {Bauer}, \&
  {Marcout}}]{Ochsenbein2000}
{Ochsenbein}, F., {Bauer}, P., \& {Marcout}, J. 2000, \aaps, 143, 23

\bibitem[Odenwald(1989)]{Odenwald1989} Odenwald, S.~F.\ 1989, \aj, 
97, 801 

\bibitem[{{Ogura} {et~al.}(2002){Ogura}, {Sugitani}, \& {Pickles}}]{Ogura2002}
{Ogura}, K., {Sugitani}, K., \& {Pickles}, A. 2002, \aj, 123, 2597

\bibitem[{{Ojha} {et~al.}(2006){Ojha}, {Ghosh}, {Tej}, {Verma}, {Vig},
  {Anupama}, {Sahu}, {Parihar}, {Bhatt}, {Prabhu}, {Maheswar}, {Bhatt},
  {Anandarao}, \& {Venkataraman}}]{Ojha2006}
{Ojha}, D.~K. {et~al.} 2006, \mnras, 368, 825

\bibitem[{{Oke} {et~al.}(1995){Oke}, {Cohen}, {Carr}, {Cromer}, {Dingizian},
  {Harris}, {Labrecque}, {Lucinio}, {Schaal}, {Epps}, \& {Miller}}]{Oke1995}
{Oke}, J.~B. {et~al.} 1995, \pasp, 107, 375

\bibitem[{{Oke} \& {Gunn}(1982)}]{Oke1982}
{Oke}, J.~B., \& {Gunn}, J.~E. 1982, \pasp, 94, 586

\bibitem[{{Peneva} {et~al.}(2010){Peneva}, {Semkov}, {Munari}, \&
  {Birkle}}]{Peneva2010}
{Peneva}, S.~P., {Semkov}, E.~H., {Munari}, U., \& {Birkle}, K. 2010, \aap,
  515, A24
  
  \bibitem[Perley et al.(2010)]{Perley2010} Perley, D.~A., et al.\ 
2010, \mnras, 406, 2473 

\bibitem[Phillips 
\& Davis(1987)]{Phillips1987} Phillips, J.~G., \& Davis, S.~P.\ 1987, \pasp, 99, 839 

\bibitem[{{Rau} {et~al.}(2009){Rau}, {Kulkarni}, {Law}, {Bloom}, {Ciardi},
  {Djorgovski}, {Fox}, {Gal-Yam}, {Grillmair}, {Kasliwal}, {Nugent}, {Ofek},
  {Quimby}, {Reach}, {Shara}, {Bildsten}, {Cenko}, {Drake}, {Filippenko},
  {Helfand}, {Helou}, {Howell}, {Poznanski}, \& {Sullivan}}]{Rau2009}
{Rau}, A. {et~al.} 2009, \pasp, 121, 1334

\bibitem[{{Rayner} {et~al.}(2003){Rayner}, {Toomey}, {Onaka}, {Denault},
  {Stahlberger}, {Vacca}, {Cushing}, \& {Wang}}]{Rayner2003}
{Rayner}, J.~T., {Toomey}, D.~W., {Onaka}, P.~M., {Denault}, A.~J.,
  {Stahlberger}, W.~E., {Vacca}, W.~D., {Cushing}, M.~C., \& {Wang}, S. 2003,
  \pasp, 115, 362

\bibitem[Rebull et al.(2010)]{Rebull2010} Rebull, L.~M., et al.\ 
2010, \apjs, 186, 259 

\bibitem[{{Reipurth} \& {Aspin}(2004{\natexlab{a}})}]{Reipurth2004}
{Reipurth}, B., \& {Aspin}, C. 2004{\natexlab{a}}, \apjl, 606, L119

\bibitem[{{Reipurth} \& {Aspin}(2004{\natexlab{b}})}]{Reipurth2004b}
---. 2004{\natexlab{b}}, \apjl, 608, L65

\bibitem[{{Reipurth} {et~al.}(2007){Reipurth}, {Aspin}, {Beck}, {Brogan},
  {Connelley}, \& {Herbig}}]{Reipurth2007}
{Reipurth}, B., {Aspin}, C., {Beck}, T., {Brogan}, C., {Connelley}, M.~S., \&
  {Herbig}, G.~H. 2007, \aj, 133, 1000

\bibitem[{{Sharon} {et~al.}(2010){Sharon}, {Hillenbrand}, {Fischer}, \&
  {Edwards}}]{Sharon2010}
{Sharon}, C., {Hillenbrand}, L., {Fischer}, W., \& {Edwards}, S. 2010, \aj,
  139, 646

\bibitem[Sharp 
\& Burrows(2007)]{Sharp2007} Sharp, C.~M., \& Burrows, A.\ 2007, \apjs, 168, 140 

\bibitem[{{Skrutskie} {et~al.}(2006){Skrutskie}, {Cutri}, {Stiening},
  {Weinberg}, {Schneider}, {Carpenter}, {Beichman}, {Capps}, {Chester},
  {Elias}, {Huchra}, {Liebert}, {Lonsdale}, {Monet}, {Price}, {Seitzer},
  {Jarrett}, {Kirkpatrick}, {Gizis}, {Howard}, {Evans}, {Fowler}, {Fullmer},
  {Hurt}, {Light}, {Kopan}, {Marsh}, {McCallon}, {Tam}, {Van Dyk}, \&
  {Wheelock}}]{Skrutskie2006}
{Skrutskie}, M.~F. {et~al.} 2006, \aj, 131, 1163

\bibitem[{{Steidel} {et~al.}(2004){Steidel}, {Shapley}, {Pettini},
  {Adelberger}, {Erb}, {Reddy}, \& {Hunt}}]{Steidel2004}
{Steidel}, C.~C., {Shapley}, A.~E., {Pettini}, M., {Adelberger}, K.~L., {Erb},
  D.~K., {Reddy}, N.~A., \& {Hunt}, M.~P. 2004, \apj, 604, 534

\bibitem[{{Storey} \& {Hummer}(1995)}]{Storey1995}
{Storey}, P.~J., \& {Hummer}, D.~G. 1995, \mnras, 272, 41

\bibitem[{{Szeifert} {et~al.}(2010){Szeifert}, {Hubrig}, {Sch{\"o}ller},
  {Sch{\"u}tz}, {Stelzer}, \& {Mikul{\'a}{\v s}ek}}]{Szeifert2010}
{Szeifert}, T., {Hubrig}, S., {Sch{\"o}ller}, M., {Sch{\"u}tz}, O., {Stelzer},
  B., \& {Mikul{\'a}{\v s}ek}, Z. 2010, \aap, 509, L7+

\bibitem[{{Vacca} {et~al.}(2003){Vacca}, {Cushing}, \& {Rayner}}]{Vacca2003}
{Vacca}, W.~D., {Cushing}, M.~C., \& {Rayner}, J.~T. 2003, \pasp, 115, 389

\bibitem[{{Vacca} {et~al.}(2004){Vacca}, {Cushing}, \& {Simon}}]{Vacca2004}
{Vacca}, W.~D., {Cushing}, M.~C., \& {Simon}, T. 2004, \apjl, 609, L29

\bibitem[{{van Dokkum}(2001)}]{v01}
{van Dokkum}, P.~G. 2001, \pasp, 113, 1420

\bibitem[Voges et al.(1999)]{Voges1999} Voges, W., et al.\ 1999, \aap, 349, 389 

\bibitem[{{Vorobyov} \& {Basu}(2005)}]{Vorobyov2005}
{Vorobyov}, E.~I., \& {Basu}, S. 2005, \apjl, 633, L137

\bibitem[{{Wade} \& {Horne}(1988)}]{wh88}
{Wade}, R.~A., \& {Horne}, K. 1988, \apj, 324, 411

\bibitem[{{Walter} {et~al.}(2004){Walter}, {Stringfellow}, {Sherry}, \&
  {Field-Pollatou}}]{Walter2004}
{Walter}, F.~M., {Stringfellow}, G.~S., {Sherry}, W.~H., \& {Field-Pollatou},
  A. 2004, \aj, 128, 1872

\bibitem[{{Welin}(1973)}]{Welin1973}
{Welin}, G. 1973, \aaps, 9, 183

\bibitem[{{Whelan} {et~al.}(2010){Whelan}, {Dougados}, {Perrin}, {Bonnefoy},
  {Bains}, {Redman}, {Ray}, {Bouy}, {Benisty}, {Bouvier}, {Chauvin}, {Garcia},
  {Grankvin}, \& {Malbet}}]{Whelan2010}
{Whelan}, E.~T. {et~al.} 2010, \apjl, 720, L119

\bibitem[{{White} \& {Hillenbrand}(2004)}]{White2004}
{White}, R.~J., \& {Hillenbrand}, L.~A. 2004, \apj, 616, 998

\bibitem[{{Wilson} {et~al.}(2004){Wilson}, {Henderson}, {Herter}, {Matthews},
  {Skrutskie}, {Adams}, {Moon}, {Smith}, {Gautier}, {Ressler}, {Soifer}, {Lin},
  {Howard}, {LaMarr}, {Stolberg}, \& {Zink}}]{Wilson2004}
{Wilson}, J.~C. {et~al.} 2004, in Presented at the Society of Photo-Optical
  Instrumentation Engineers (SPIE) Conference, Vol. 5492, Society of
  Photo-Optical Instrumentation Engineers (SPIE) Conference Series, ed.
  {A.~F.~M.~Moorwood \& M.~Iye}, 1295

\bibitem[{{Zacharias} {et~al.}(2005){Zacharias}, {Monet}, {Levine}, {Urban},
  {Gaume}, \& {Wycoff}}]{Zacharias2005}
{Zacharias}, N., {Monet}, D.~G., {Levine}, S.~E., {Urban}, S.~E., {Gaume}, R.,
  \& {Wycoff}, G.~L. 2005, VizieR Online Data Catalog, 1297, 0

\bibitem[{{Zhu} {et~al.}(2009){Zhu}, {Hartmann}, {Gammie}, \&
  {McKinney}}]{Zhu2009}
{Zhu}, Z., {Hartmann}, L., {Gammie}, C., \& {McKinney}, J.~C. 2009, \apj, 701,
  620

\bibitem[{{Zickgraf} {et~al.}(1989){Zickgraf}, {Wolf}, {Stahl}, \&
  {Humphreys}}]{Zickgraf1989}
{Zickgraf}, F., {Wolf}, B., {Stahl}, O., \& {Humphreys}, R.~M. 1989, \aap, 220,
  206

\end{thebibliography}

\setlength{\baselineskip}{1.667\baselineskip}

\clearpage

\LongTables

\begin{deluxetable}{lcc}
\tablewidth{0pt}
\tabletypesize{\tiny}
\tablecaption{P48 $R$-Band Photometry of PTF10nvg \label{tab:P48}}
\tablehead{
\colhead{Epoch (JD }  & 
\colhead{} & 
\colhead{1 $\sigma$} \\ 
\colhead{$-$ 2,400,000)}  & 
\colhead{mag} & 
\colhead{error} }
\startdata
    55056.72052 & 17.791 &  0.058 \\
    55056.79538 & 17.769 &  0.054 \\
    55059.76152 & 17.335 &  0.038 \\
    55059.85698 & 17.357 &  0.042 \\
    55061.81557 & 17.500 &  0.043 \\
    55061.88339 & 17.527 &  0.048 \\
    55062.82658 & 17.650 &  0.051 \\
    55062.89059 & 17.645 &  0.053 \\
    55064.78602 & 17.838 &  0.058 \\
    55064.83105 & 17.827 &  0.059 \\
    55067.81776 & 17.898 &  0.206 \\
    55067.89291 & 18.045 &  0.073 \\
    55080.79022 & 19.209 & 99.000 \\
    55080.83392 & 19.229 & 99.000 \\
    55080.86080 & 19.133 & 99.000 \\
    55088.81558 & 18.985 &  0.147 \\
    55093.81772 & 19.095 &  0.141 \\
    55094.75246 & 19.275 & 99.000 \\
    55107.72160 & 19.166 & 99.000 \\
    55107.76558 & 19.207 & 99.000 \\
    55123.72473 & 19.103 & 99.000 \\
    55302.97917 & 15.475 &  0.014 \\
    55303.91102 & 15.490 &  0.013 \\
    55303.95934 & 15.501 &  0.011 \\
    55310.86996 & 15.361 &  0.012 \\
    55310.91303 & 15.340 &  0.008 \\
    55316.87539 & 15.450 &  0.013 \\
    55316.91979 & 15.443 &  0.012 \\
    55322.85902 & 15.139 &  0.011 \\
    55322.90278 & 15.158 &  0.010 \\
    55328.90851 & 14.729 &  0.007 \\
    55328.95302 & 14.745 &  0.006 \\
    55335.93323 & 14.585 &  0.006 \\
    55335.97744 & 14.562 &  0.006 \\
    55340.92922 & 14.691 &  0.006 \\
    55340.97352 & 14.653 &  0.007 \\
    55345.96956 & 14.761 &  0.008 \\
    55346.76993 & 14.772 &  0.007 \\
    55346.81387 & 14.777 &  0.007 \\
    55351.77118 & 14.779 &  0.008 \\
    55351.81455 & 14.770 &  0.007 \\
    55356.80886 & 14.499 &  0.005 \\
    55356.85329 & 14.465 &  0.005 \\
    55361.81103 & 15.488 &  0.009 \\
    55361.85461 & 15.462 &  0.010 \\
    55366.81772 & 15.475 &  0.011 \\
    55366.86154 & 15.380 &  0.015 \\
    55371.83990 & 15.579 &  0.009 \\
    55371.88355 & 15.597 &  0.009 \\
    55376.87708 & 16.136 &  0.016 \\
    55376.92012 & 16.122 &  0.015 \\
    55380.96681 & 16.316 &  0.019 \\
    55381.90493 & 15.992 &  0.013 \\
    55381.95505 & 15.986 &  0.013 \\
    55386.89716 & 15.962 &  0.013 \\
    55386.94417 & 15.955 &  0.013 \\
    55391.89965 & 15.418 &  0.015 \\
    55391.94368 & 15.549 &  0.009 \\
    55396.91148 & 15.973 &  0.012 \\
    55396.96554 & 15.879 &  0.019 \\
    55401.80397 & 15.828 &  0.011 \\
    55401.84827 & 15.801 &  0.012 \\
    55407.79447 & 15.199 &  0.013 \\
    55407.84208 & 15.225 &  0.012 \\
    55410.79516 & 14.874 &  0.005 \\
    55410.84040 & 14.843 &  0.008 \\
    55413.79185 & 14.244 &  0.004 \\
    55413.83559 & 14.296 &  0.004 \\
    55416.78713 & 14.426 &  0.005 \\
    55416.83032 & 14.434 &  0.005 \\
    55419.78179 & 14.358 &  0.004 \\
    55419.82614 & 14.340 &  0.005 \\
    55422.94676 & 14.440 &  0.005 \\
    55422.99104 & 14.402 &  0.005 \\
    55426.66052 & 14.333 &  0.005 \\
    55429.67674 & 13.851 &  0.002 \\
    55429.72018 & 13.847 &  0.003 \\
    55432.74426 & 13.544 &  0.002 \\
    55432.78742 & 13.547 &  0.002 \\
    55435.75871 & 13.872 &  0.219 \\
    55435.80280 & 13.481 &  0.002 \\
    55438.83025 & 13.494 &  0.002 \\
    55438.91326 & 13.519 &  0.002 \\
    55441.86536 & 13.483 &  0.002 \\
    55441.93003 & 13.447 &  0.002 \\
    55444.88985 & 13.600 &  0.002 \\
    55444.93396 & 13.630 &  0.002 \\
    55448.64422 & 14.133 &  0.004 \\
    55448.68887 & 14.163 &  0.004 \\
    55451.65597 & 14.377 &  0.004 \\
    55451.70022 & 14.357 &  0.004 \\
    55454.72106 & 14.991 &  0.010 \\
    55454.76510 & 15.069 &  0.007 \\
    55457.71681 & 14.692 &  0.007 \\
    55457.76084 & 14.720 &  0.006 \\
    55460.83892 & 15.044 &  0.007 \\
    55460.88752 & 15.009 &  0.007 \\
    55463.84071 & 15.050 &  0.010 \\
    55463.88587 & 15.091 &  0.008 \\
    55466.82553 & 15.143 &  0.008 \\
    55466.87714 & 15.139 &  0.009 \\
    55471.60936 & 15.066 &  0.007 \\
    55471.68838 & 15.040 &  0.010 \\
    55477.69095 & 14.578 &  0.004 \\
    55477.73424 & 14.554 &  0.006 \\
    55480.74010 & 14.701 &  0.007 \\
    55480.78351 & 14.696 &  0.006 \\
    55484.76268 & 14.854 &  0.008 \\
    55485.61926 & 14.769 &  0.007 \\
    55485.66346 & 14.786 &  0.006 \\
    55497.59777 & 15.392 &  0.010
    \enddata
\end{deluxetable}

\begin{deluxetable}{lccc}
\tablewidth{0pt}
\tabletypesize{\tiny}
\tablecaption{P60 Optical Photometry of PTF10nvg \label{tab:P60}}
\tablehead{
\colhead{Epoch (JD}  & 
\colhead{} & 
\colhead{} & 
\colhead{1 $\sigma$}  \\
\colhead{$-$ 2,400,000)}  & 
\colhead{filter} & 
\colhead{mag} & 
\colhead{error} }
\startdata
   55409.353  & $r $ & 15.487 &  0.170 \\
   55409.355  & $z $ & 13.274 &  0.254 \\
   55410.177  & $z $ & 13.134 &  0.332 \\
   55411.234  & $z $ & 13.096 &  0.189 \\
   55412.206  & $i $ & 14.208 &  0.025 \\
   55412.207  & $z $ & 12.830 &  0.225 \\
   55413.326  & $r $ & 14.776 &  0.150 \\
   55413.327  & $z $ & 12.772 &  0.258 \\
   55414.255  & $z $ & 12.864 &  0.266 \\
   55415.280  & $z $ & 12.908 &  0.230 \\
   55416.279  & $i $ & 14.218 &  0.043 \\
   55416.280  & $z $ & 12.935 &  0.196 \\
   55417.244  & $r $ & 14.920 &  0.159 \\
   55417.245  & $z $ & 12.859 &  0.285 \\
   55418.377  & $z $ & 12.947 &  0.240 \\
   55419.316  & $z $ & 12.856 &  0.215 \\
   55420.184  & $i $ & 14.105 &  0.040 \\
   55420.185  & $z $ & 12.795 &  0.238 \\
   55421.267  & $r $ & 15.059 &  0.162 \\
   55421.268  & $z $ & 13.026 &  0.267 \\
   55423.183  & $z $ & 12.994 &  0.188 \\
   55424.408  & $i $ & 14.241 &  0.034 \\
   55424.410  & $z $ & 12.896 &  0.257 \\
   55433.350  & $i $ & 13.360 &  0.113 \\
   55433.351  & $r $ & 13.877 &  0.191 \\
   55433.352  & $z $ & 12.661 &  0.238 \\
   55434.286  & $z $ & 11.860 &  0.546 \\
   55436.256  & $z $ & 12.001 &  0.226 \\
   55438.220  & $i $ & 13.285 &  0.226 \\
   55438.298  & $r $ & 13.969 &  0.121 \\
   55438.299  & $z $ & 12.089 &  0.153 \\
   55442.213  & $i $ & 13.246 &  0.002 \\
   55442.214  & $r $ & 13.899 &  0.002 \\
   55442.214  & $z $ & 12.598 &  0.001 \\
   55443.218  & $i $ & 13.219 &  0.018 \\
   55443.220  & $r $ & 13.798 &  0.160 \\
   55443.225  & $z $ & 11.963 &  0.250 \\
   55444.239  & $i $ & 13.402 &  0.029 \\
   55444.240  & $r $ & 13.978 &  0.186 \\
   55444.241  & $z $ & 12.172 &  0.233 \\
   55445.261  & $i $ & 13.572 &  0.020 \\
   55445.265  & $r $ & 14.176 &  0.181 \\
   55445.277  & $z $ & 12.336 &  0.231 \\
   55446.280  & $i $ & 13.587 &  0.020 \\
   55446.281  & $r $ & 14.194 &  0.181 \\
   55446.282  & $z $ & 12.311 &  0.230 \\
   55447.236  & $i $ & 13.875 &  0.029 \\
   55447.237  & $z $ & 12.576 &  0.219 \\
   55448.164  & $z $ & 12.693 &  0.196 \\
   55449.372  & $z $ & 12.746 &  0.137 \\
   55450.161  & $r $ & 14.513 &  0.161 \\
   55450.162  & $z $ & 12.537 &  0.282 \\
   55451.166  & $r $ & 14.863 &  0.129 \\
   55452.355  & $z $ & 13.002 &  0.225 \\
   55458.340  & $r $ & 15.241 &  0.159 \\
   55458.341  & $z $ & 13.158 &  0.309 \\
   55459.311  & $z $ & 13.334 &  0.200 \\
   55460.228  & $z $ & 13.455 &  0.079 \\
   55461.197  & $z $ & 13.382 &  0.144 \\
   55462.270  & $r $ & 15.534 &  0.181 \\
   55462.271  & $z $ & 13.463 &  0.122 \\
   55463.187  & $z $ & 13.463 &  0.116 \\
   55469.183  & $r $ & 15.763 &  0.190 \\
   55469.184  & $z $ & 13.611 &  0.179 \\
   55470.362  & $z $ & 13.576 &  0.160 \\
   55473.308  & $r $ & 15.486 &  0.139 \\
   55473.309  & $z $ & 13.421 &  0.138 \\
   55477.322  & $r $ & 15.059 &  0.161 \\
   55477.324  & $z $ & 13.042 &  0.166 \\
   55478.146  & $z $ & 13.129 &  0.148 \\
   55479.117  & $z $ & 13.047 &  0.075 \\
   55480.171  & $z $ & 12.585 &  0.547 \\
   55481.147  & $r $ & 15.335 &  0.089 \\
   55481.148  & $z $ & 13.250 &  0.099 \\
   55482.108  & $z $ & 13.317 &  0.160 \\
   55483.128  & $z $ & 13.356 &  0.080 \\
   55484.182  & $z $ & 13.360 &  0.179 \\
   55485.109  & $r $ & 15.283 &  0.146 \\
   55485.109  & $z $ & 13.321 &  0.177 \\
   55496.280  & $r $ & 16.102 &  0.233 \\
   55496.281  & $z $ & 14.078 &  0.228 \\
   55497.130  & $z $ & 13.919 &  0.174 \\
   55499.122  & $z $ & 13.969 &  0.103
\enddata
\end{deluxetable}

\clearpage

\begin{deluxetable}{lcccccc}
\tablewidth{0pt}
\tabletypesize{\tiny}
\tablecaption{NIR Photometry of PTF10nvg \label{tab:Paritelmags}}
\tablehead{
\colhead{Epoch (JD)}  & 
\colhead{$J$ mag} & 
\colhead{$J$ err} & 
\colhead{$H$ mag} & 
\colhead{$H$ err} & 
\colhead{$K$ mag} & 
\colhead{$K$ err} }
\startdata
2451823.66652 & $>$16.35  & \nodata & $>$15.38  & \nodata & $>$14.80 & \nodata \\
2455387.87633  &  12.22    &   0.07    &   10.26    &   0.02    &   8.45    &   0.01 \\
2455392.79441   &     12.23    &   0.01    &   10.38    &   0.06    &   8.45    &   0.02 \\
2455450.69076    &   10.88    &   0.02   & 9.39   & 0.01    &   8.06    &   0.01\\
2455463.65116    &   11.48    &   0.03    &   9.86    &   0.01    &   8.39    &   0.01 \\
2455466.62568    &   11.62    &   0.03    &   10.02    &   0.01    &   8.38    &   0.01 \\
\enddata
\end{deluxetable}

\begin{deluxetable}{lccccc}
\tablewidth{0pt}
\tabletypesize{\tiny}
\tablecaption{Pre-Outburst Photometry of PTF10nvg \label{tab:precursor_phot}}
\tablehead{
\colhead{Wavelength}  & 
\colhead{Flux } & 
\colhead{Flux } & 
\colhead{} & 
\colhead{} & 
\colhead{} \\
\colhead{$\mu$m}  & 
\colhead{Density (Jy)} & 
\colhead{Error (1$\sigma$) } & 
\colhead{Observatory} & 
\colhead{Epoch} & 
\colhead{Reference} }
\startdata
12 & 3.39 & 0.33 & IRAS & 1983 & \citet{Odenwald1989} \\
25 & 6.59 & 0.66 & IRAS & 1983 & \citet{Odenwald1989} \\
60 & 27.89 & 1.0 & IRAS & 1983 & \citet{Odenwald1989} \\
100 & 57.35 & 1.0 & IRAS & 1983 & \citet{Odenwald1989} \\
8.28 & 1.6 & 0.66 & MSX & 1996-7 & \citet{Egan1999} \\
12.13 & 2.49 & 0.137 &MSX & 1996-7 & \citet{Egan1999} \\
14.65 & 3.045 & 0.189 &MSX & 1996-7 & \citet{Egan1999} \\
21.34 & 3.14 & 0.201 &MSX & 1996-7 & \citet{Egan1999} \\
3.6 & 0.083820 &  0.0041974110 & {\it Spitzer}/IRAC & 2006-7 & \citet{Guieu2009} \\
4.5 & 0.231500  &  0.0011582866 & {\it Spitzer}/IRAC & 2006-7 & \citet{Guieu2009} \\ 
5.8 & 0.609100  &  0.0030475698 & {\it Spitzer}/IRAC & 2006-7 & \citet{Guieu2009} \\
8.0 & 1.171000  &  0.0058589830 & {\it Spitzer}/IRAC & 2006-7 & \citet{Guieu2009} \\
24. & 2.222000  &  0.088946621 & {\it Spitzer}/MIPS & 2006-7 & Rebull et al., in prep. \\
70. & 6.732000  &  \nodata  & {\it Spitzer}/MIPS & 2006-7 & Rebull et al., in prep. \\
9.     &       1.96   &  0.2  & AKARI & 2006-7 & \citet{Ishihara2010} \\
15.    &       3.875 &   0.457 & AKARI & 2006-7 & \citet{Ishihara2010}
\enddata
\end{deluxetable}

\clearpage

\begin{landscape}

\begin{deluxetable}{lccccccc}
\tablewidth{0pt}
\tabletypesize{\tiny}
\tablecaption{Optical Line Strengths \label{tab:opt_lines}}
\tablehead{
  \colhead{} &
  \colhead{Wavelength} &
 \colhead{LRIS} &
  \colhead{LRIS flux} &
  \colhead{Kast} &
  \colhead{Kast flux} &
\colhead{DoubleSpec} &
\colhead{DoubleSpec} \\
 \colhead{Line} &
  \colhead{(\AA\ )} &
 \colhead{EqW (\AA\ )} &
  \colhead{(10$^{-15}$ erg cm$^{-2}$ s$^{-1}$)} &
  \colhead{EqW (\AA\ )} &
  \colhead{(10$^{-15}$ erg cm$^{-2}$ s$^{-1}$)} &
 \colhead{EqW (\AA\ )} &
\colhead{flux (10$^{-15}$  erg cm$^{-2}$ s$^{-1}$)} }\startdata
\\
H$\eta$ & 3835.38 & 3.68 & 0.087 & 4.15 & 0.21 & \nodata & \nodata \\
H$\zeta$ & 3889.05 & 2.57 & 0.14 & 6.57 & 0.30 & \nodata & \nodata \\
Ca II K (abs) & 3933.66 & 1.41 & 0.056 & 4.01 & 0.25 & \nodata & \nodata \\
Ca II K (em) & 3933.66 & -5.65 & 0.24 & -10.57 & 0.449 & \nodata & \nodata \\
H$\epsilon$ + Ca II H& 3970.07 & 4.93 & 0.21 & 7.04 & 0.366 & \nodata & \nodata \\
H$\delta$ & 4101.74 & 2.17 & 0.12 & 5.34 & 0.306 & \nodata & \nodata \\
H$\gamma$ & 4340.47 & 4.00 & 0.33 & 7.13 & 0.57 & \nodata & \nodata \\
H$\beta$ (abs) & 4861.33 & 2.88 & 0.46 & 3.92 & 0.55 & \nodata & \nodata \\
H$\beta$ (em) & 4861.33 & -1.7 & 0.27 & -0.3 & 0.04 & \nodata & \nodata \\
Na D & 5889.95+5895.92 & 9.637 & 4.35 & 7.82 & 2.85 & \nodata & \nodata \\
Fe I & 6191.56 & -0.61 & 0.368 & -1.82 & 0.89 & \nodata & \nodata \\
$[$O I$]$ & 6300.30 & -9.14 & 5.48 & -7.91 & 3.99 & \nodata & \nodata \\
$[$O I$]$ & 6363.78 & -3.8 & 2.245 & -2.95 & 1.41 & -1.5 & 2.22 \\
Fe II & 6432.68 & -2.11 & 1.33 & -3.38 & 1.59 & -1.68 & 2.73 \\
Fe I & 6495.74 & -1.158 & 0.80 & -2.27 & 1.17 & -1.78 & 2.97 \\
Fe II & 6516.08 & -2.13 & 1.5 & -2.84 & 1.49 & -2.92 & 4.89 \\
H$\alpha$ & 6562.85 & -39.9 & 31.5 & -56.5 & 31.33 & -24.7 & 48.6 \\
$[$N II$]$ & 6592 & -0.86 & 0.70 & -0.18 & 0.13 & -1.15 & 2.23 \\
$[$S II$]$ & 6716.44 & -0.34 & 0.29 & 1.26 & 0.89 & \nodata & \nodata \\
$[$S II$]$ & 6730.82 & -0.80 &0.67 & 0.82 & 0.58 & \nodata & \nodata \\
Ca I & 7148.15 & -1.83 & 2.04 & -1.76 & 1.64 & \nodata & \nodata \\
$[$Fe II$]$ & 7172.00 & -0.59 & 0.66 & 0.58 & 0.56 & \nodata & \nodata \\
$[$Ca II$]$ & 7291.47 & -1.08 & 1.23 & -1.63 & 1.46  & -1.14 & 3.47 \\
$[$Ca II$]$ & 7323.89 & -1.02 & 1.13 & -1.12 & 0.97 & -1.19 & 3.6 \\
K I & 7664.91 & 3.02 & 4.22 & 2.24 & 2.66 & 3.16 & 11.1 \\
K I & 7698.96 & 2.32 & 3.32 & 2.86 & 3.80 & 1.54 & 5.74 \\
O I & 7773 & 2.26 & 3.42 & 2.43 & 3.35 & 1.77 & 7.32 \\
O I & 8446.36 & -1.52 & 2.86 & -2.47 & 3.98 & -2.16 & 9.81 \\
Ca II & 8498.02 & -16.8 & 32.8 & -23.46 & 35.57 & -19.0 & 85.6 \\
Ca II & 8542.09 & -19.1 & 35.9 & -24.24 & 37.63 & -16.2 & 69.3 \\
$[$Fe II$]$ & 8616.952 & -1.14 & 2.1 & -0.73 & 1.11 & \nodata & \nodata \\
Ca II & 8662.14 & -15.4 & 29.0 & -24.88 & 36.7 & -17.55 & 60.8 \\
H I (Pa 11) & 8863.4 & -1.47 & 2.64 & -3.14 & 4.14 & \nodata & \nodata \\
H I (Pa 10) & 9015.6 & -1.58 & 2.98 & -2.15 & 2.80 & \nodata & \nodata \\
H I (Pa 9) & 9229.7 & -1.61 & 3.46 & -2.73 & 4.26 & \nodata & \nodata \\
H I (Pa 8) & 9545.97 & -1.44 & 3.36 & -3.59 & 5.19 & \nodata & \nodata \\
$[$Fe I$]$ & 9998.33 & -2.28 & 5.21 & \nodata & \nodata & \nodata & \nodata \\
H I (Pa 7) & 10049.4 & -1.82 & 4.89 & -6.35 & 9.3 & \nodata & \nodata \\
\enddata
\end{deluxetable}

\clearpage
\end{landscape}

\begin{deluxetable}{lccccc}
\tablewidth{0pt}
\tabletypesize{\tiny}
\tablecaption{Near--Infrared Line Strengths \label{tab:nir_lines}}
\tablehead{
  \colhead{} &
  \colhead{Wavelength} &
 \colhead{SpeX} &
  \colhead{SpeX flux} &
 \colhead{TripleSpec} &
  \colhead{TripleSpec flux}\\
 \colhead{Line} &
  \colhead{($\mu$m)} &
 \colhead{EqW (\AA\ )} &
  \colhead{(10$^{-15}$ erg cm$^{-2}$ s$^{-1}$)} &
 \colhead{EqW (\AA\ )} &
  \colhead{(10$^{-15}$ erg cm$^{-2}$ s$^{-1}$)} 
}\startdata
\\
Ca II & 0.8662 &  -12.91 & -26.4 & \nodata & \nodata \\
Ca II & 0.8542 & -13.49 & -28.5 & \nodata & \nodata \\
Ca II & 0.8499 & -13.97 & -30.73 & \nodata & \nodata \\
H I 7-3 & 1.0049 & -0.698 & -1.706 & -4.04 & -7.10 \\
H I 6-3 & 1.0938 & -3.14 & -8.625 & -16.01 & -19.4 \\
O I & 1.1287 & -4.211 & -12.397 & -13.01 & 10.06 \\
Fe I & 1.1595 & -0.733 & -2.293 & -1.15 & -2.29 \\
Fe I & 1.161 & -1.294 & -4.152 & -3.30 & -6.83 \\
K I & 1.169 & -1.449 & -4.539 & -3.13832 & -6.80 \\
Fe I & 1.179 & -0.807 & -2.665 & -1.45 & -3.55 \\
Mg I & 1.1833 & -2.092 & -6.881 & -2.64 & -6.16 \\
Fe I & 1.1885 & -3.018 & -10.106 & -5.26 & -12.47 \\
Fe I & 1.1974 & -1.847 & -6.444 & -3.48 & -8.89 \\
H I 5-3 & 1.2818 & -7.766 & -31.026 & -16.5 & -50.4 \\
Mg I & 1.504 & -3.777 & -23.54 & -5.39 & -29.7 \\
Si I & 1.589 & -2.372 & -16.094 & -2.47 & -15.3 \\
Mg I & 1.711 & -0.587 & -4.73 & -0.73 & -5.50 \\
H$_2$ (1-0 S(1)) & 2.121 & -0.848 & -10.12 & -1.35 & -16.9 \\
H I 7-4 (Br $\gamma$) & 2.1657 & -1.186 & -14.40 & -2.24 & -28.58 \\
Na I & 2.2075 & -0.768 & -9.53 & -0.957 & -12.49 \\
H$_2$ (2-1 S(1)) & 2.247 & -0.199 & -2.412 & -0.41 & -5.52 \\
Ca I & 2.82 & -0.503 & -6.5 & -1.07 & -14.9 \\
CO & 2.923 & -18.8 & -249.9 & -29.3 & -423.9 \\
\enddata
\end{deluxetable}

\end{document}